\documentclass[11pt,a4paper]{article}
\usepackage{jheppub_kim}
 \usepackage{slashed}

\usepackage{longtable}

\usepackage{epsfig}
\usepackage{amsmath}
\usepackage{graphicx}
 \usepackage{graphics}
\usepackage[hang,nooneline,scriptsize]{subfigure}

\usepackage{array}

\begin{document}
\title{\Large   Smeared Mass Source Wormholes in Modified $f(R)$ Gravity with the Lorentzian Density Distribution Function}

\author[a]{J. Sadeghi,}
\author[b]{B. Pourhassan,}
\author[c]{S. Noori Gashti,}
\author[d,e,b,*]{and S. Upadhyay 
\note{Presently, Visiting Associate   at Inter-University Centre for Astronomy and Astrophysics (IUCAA),
Pune, Maharashtra 411007, India.}}

\affiliation[a] {Department of Physics, Faculty of Basic Sciences, University of Mazandaran P. O. Box
47416-95447, Babolsar, Iran.}
\affiliation[a] {Canadian Quantum Research Center 204-3002 32 Avenue Vernon, British Columbia V1T 2L7 Canada.}

\affiliation[b] {School of Physics, Damghan University, P. O. Box 3671641167, Damghan, Iran.}

\affiliation[c] {Department of Physics, University of Mazandaran P. O. Box 47416-95447, Babolsar, Iran.}

\affiliation[d] {Department of Physics, K.L.S. College,  Nawada, Bihar  805110, India.}
\affiliation[e] {Department of Physics, Magadh University, Bodh Gaya, Bihar 824234, India}

\emailAdd{pouriya@ipm.ir}

\emailAdd{b.pourhassan@du.ac.ir}

\emailAdd{saeed.noorigashti@stu.umz.ac.ir}

\emailAdd{sudhakerupadhyay@gmail.com; sudhaker@associates.iucaa.in}

\abstract{Wormholes are speculative structures linking disparate space-time points. Their geometry can be obtained by solving Einstein equations with tolerating the violation of null energy conditions. Recently, many researchers have studied different wormholes according to different criteria, and they achieved remarkable results. In this paper, we investigate a series of exact solutions of the static wormhole with smeared mass source geometry in modified $f(R)$ gravity theories. In fact, we consider the Lorentzian density distribution which is coming from a particle-like source. To be more specific, the modified gravity models we consider here are some power laws. We compute resulting solutions according to the wormhole field equations. We also specify parameters such as the radial pressure and transverse pressure as well as various energy conditions such as null energy conditions, weak energy conditions and strong energy conditions. Finally, by plotting some figures, in addition to identifying the wormhole throat, we describe the results of either the violation or the satisfaction of the energy conditions completely.}



\maketitle
 \section{Introduction}
The concept of wormhole, in general, was first introduced by Flamm in 1916 \cite{1}. However, a few years later, Einstein and Rosen explored a similar structure with concept of wormhole in 1935, and they introduced a similar geometric structure which is called the Einstein-Rosen Bridge \cite{2}. In fact, recently, wormholes have been evaluated by many researchers according to different criteria. In following, we will introduce a number of these studies. Actually, wormhole is a speculative structure linking disparate space-time points and is based on a particular solution of the Einstein field equations. A wormhole can be visualized as a tunnel with two ends at separate points in space-time, i.e., different locations, various points in time, or both\cite{mult}). Geometry of static symmetrical spherical wormholes consists of a two-state tunnel is usually considered in the literature which is called different names such as throat, tube or handle. In general, it is possible for a two-way journey if both distinct points belong to the same space-time. In the last few years, wormholes have been studied from different points of view \cite{mult2,mult3}.

Different types of  wormholes have been evaluated in different conditions in the literature, including cylindrical symmetry, non-static symmetries, conditions supported by cosmological constants, thin shells and static wormholes with electric charge, as well as various other types that have been explored by researchers over the past few years. However, despite all these studies, the wormhole is still a hypothetical object and no mechanism has been introduced to observe it, see Refs. \cite{3,4,5,6,7,8,9,10,11,12,13,14,15,16,17} for further reading.  Perhaps one of the most interesting and exciting types of wormholes is the traversable wormholes \cite{18,19}. In addition to the above issues, wormholes have also been studied from a cosmological point of view \cite{20}. In fact, all of these studies with different criteria to find out the truth of what is hidden in the universe and researchers beings seek to discover its facts.  Despite all these issues raised, wormholes exhibit a number of esoteric properties behaviors, such as violation of Hawking chronology protection conjecture, and in fact pursue cases that are associated with speed beyond light, or in other words, violation of causality \cite{21,22,23,24,25,26,27,28,29,30,31,32,33}. In fact, the issue we are talking about is that the energy of matter, which is somehow supported by an exotic geometry, leads to a violation of the standard energy conditions. To avoid such problems, a series of modified $f(R)$ gravity theories are used in the studies related to wormholes. In fact, relating to wormholes, different energy conditions are of great important issue that have been studied extensively in the literature. In recent decade, a large number of researchers have conducted numerous studies on the geometry of the stability of wormholes and they have obtained exact solutions for wormholes according to different field equations and various forms of equations of state and shape functions. In those case, they also discussed in detail of satisfaction or violation of energy conditions \cite{34}.

In fact, different energy conditions have been evaluated to the study of wormholes. In wormhole studies, according to the field equation, the radial pressure and transverse pressure as well as null energy conditions (NEC) are calculated and its violation or satisfaction is analyzed. Of course, we follow the same route in this work, due to the Lorentzian density distribution with a particle-like gravitational source \cite{35}. We note that here, in modified $f(R)$ gravitational theories, the stress energy tensor is replaced by the effective stress energy tensor, which is associated with a much higher curvature \cite{CJP, MNRAS}. These effective theories of gravity are commonly used in various studies, such as the problem of exotic matter in wormholes, the construction of cosmic models and the explanation of singularities. Of course, as it is clear in such theories, the geometric part of the story is modified by substituting the curvature of the scalar $R$ in action. It actually gives us the field equations that offer much more complex solutions. So, generally one can say that the geometry of a wormhole can be solved by Einstein field equations with tolerating violations of null energy conditions. As we know,  all of these issues will be possible when we have an exotic mater distribution which is not possible for physical matter distribution \cite{36}.

According to the above mentioned concepts, this paper is organized as follows.
In section \ref{s1}, we give a brief explanation of the Maurice-Thorne wormhole field equations by considering the modified $f(R)$ gravitational model \cite{IJGMMP} and having Lorentzian density distribution associated with a particle-like gravitational source. In section \ref{s2},
we present our modified $f(R)$ gravitational model as power law plus linear term ($f(R)=-\frac{R}{2}+a R^{2(1-n)}$) and
calculate energy density. Furthermore, in the subsequent subsections,
we specify the value of $n$ and derive shape function and therefore the scalar curvature, the radial pressure and tangential pressure. The condition for satisfaction or violation of different energy conditions are also presented.  In section \ref{s3},  we introduce the simple power law of the form  $f(R)=\frac{\alpha}{R^{n}}$, and follow the
similar procedure  of section \ref{s2}. For this model of $f(R)$ gravity, we have assigned different values of $n$ also.
A full comparative analysis is made corresponding to different cases. The results with future remarks are summarized  in the last section.

\section{Einstein field equations of modified gravity}\label{s1}
In this section, we shed light on  the field equations corresponding to the
Morris-Thorne wormhole, considering a modified  $f(R)$ gravity. One of the
first points we have to pay attention  is the wormhole metric, that is, the
metric of the static spherical symmetric wormhole space-time, which is given by \cite{PRD2}
\begin{equation}\label{1}
ds^{2}=-e^{2\Phi(r)}dt^{2}+\frac{dr^{2}}{1-\frac{b(r)}{r}}+r^{2}(d\theta^{2}+
\sin^{2}\theta d\phi^{2}),
\end{equation}
where $ \Phi(r) $ is the redshift function and $b(r)$ is shape function of wormholes. In fact, $ b(r_{ 0}) = r_ {0} $ is used to
determine the wormhole throat. Actually, the minimum value of $ r_ {0} $
indicates the location of the wormhole throat, which   satisfies the two most
important conditions:  $ b-\frac{b'r}{b^{2} }<1$ and $ b'(r_{0})<1 $.
In our notation prime denotes derivative with respect to $r$. According to the modified $f(R)$ gravitational model  the
energy-momentum tensor for a wormhole tensor is expressed as follows
\cite{17}
\begin{equation}\label{2.0}
T_{\nu}^{\mu}=(\rho+P_{r})u^{\mu}u_{\nu}-P_{r}g^{\mu}_{\nu}+(P_{t}-P_{r})
\eta^{\mu}\eta_{\nu}.
\end{equation}
Here, $\rho$ is the energy density, $P_r$ is the radial pressure
measured along the direction of $\eta^{\mu}$, and $P_{t}$ is the transverse
pressure measured in the direction orthogonal  to $\eta^{\mu}$.
However, $u^{\mu}$  is the  4-velocity and $\eta^{\mu} $ unit
space-like vector
in the radial direction and satisfy following relation: $u^{\mu}u_{\nu}=-
\eta^{\mu}\eta_{\nu}=1$, and $u^{\mu}\eta_{\mu}=0$.  For archived the
field equation, we  use the following gravitational equation \cite{34},
\begin{equation}\label{3.0}
G_{\mu\nu}\equiv R_{\mu\nu}-\frac{1}{2}R g_{\mu\nu}=T^{eff}_{\mu\nu},
\end{equation}
where $T^{eff}_{\mu\nu}$ refers to the effective stress-energy tensor.
Now from the above equation one can obtain,

\begin{eqnarray}
\frac{b'}{r^{2}}&=&\frac{\rho}{F}+\frac{H}{F},\nonumber\\
\frac{b}{r^{3}}&=& -\frac{P_{r}}{F}-\frac{1}{F}\left\{\left(1-\frac{b}{r}\right) \left(F''-F'\frac{b'r-b}{2r^{2}(1-\frac{b}{r})}\right)-H\right\},
\nonumber\\
\frac{b'r-b}{2r^{3}}&=&-\frac{P_{t}}{F}-\frac{1}{F}\left[\left(1-\frac{b}{r}
\right)\frac{F'}{r}-H\right].
\end{eqnarray}
Here, $F=\frac{df}{dr}$ and function $H$ is given by
\begin{equation}\label{7.0}
H(r)=\frac{1}{4}\left[FR+\left(1-\frac{b}{r}\right)\left(F''-\frac{b'r-b}
{2r^{2}(1-\frac{b}{r})}F'+\frac{2F'}{r}\right)+T\right].
\end{equation}
where $T$ is the trace of the stress-energy tensor and the curvature scalar   has following form:
\begin{equation}\label{9.0}
R=\frac{2b'}{r^{2}}.
\end{equation}
With the help of  above field equations,  one can express  the energy density $\rho$,  the radial pressure $P_{r}$ and  the tangential pressure $P_{t}$ as following:
\begin{eqnarray}
&&\rho =\frac{Fb'}{r^{2}},\label{rho} \\
&&P_{r}=-\frac{bF}{r^{3}}+\frac{F'}{2r^{2}}(b'r-b)-F''(1-\frac{b}{r}),\ \
P_{t}=-\frac{F'}{r}(1-\frac{b}{r})+\frac{F}{2r^{3}}(b-b'r).\label{p}
\end{eqnarray}
 In these equations the redshift function, $\Phi(r)$, is constant for all of the points. Actually these field equations  are the general expressions for matter threading wormhole. Considering all these equations and concepts presented here in this section  which  are very crucial for wormholes, we apply these for the case of the modified gravitational model in the next section.   In  fact, we explore the wormholes by calculating the field equations and plotting  figures and   will discuss the violation or satisfaction of different energy conditions with respect the different values of $f(R)$ component.

\section{Specific solutions  for the power law plus linear term model}\label{s2}
In this section, we illustrate the  concept of previous section for the particular $f(R)$ model. Here, we consider $f(R)$ model as \cite{37}
\begin{equation}\label{1.3}
f(R)=-\frac{R}{2}+a R^{2(1-n)},
\end{equation}
where $n$ is a constant parameter. The above equation is a particular form of deformed Starobinsky gravity. So far, different forms of it have also been studied \cite{37,38,39,40,41}. In fact, we analyze the field equations proposed in the previous section with respect to a non-commutative geometry with Lorentzian distribution. Hence, the energy density $ (\rho) $ of wormhole according to the concepts proposed and also associated with spherical symmetry and particle-like gravitational source can be expressed in the following form \cite{35},
\begin{equation}\label{1.4}
\rho=\frac{M\sqrt{\phi}}{\pi^{2}(r^{2}+\phi)^{2}},
\end{equation}
where  $\phi$ is a smeared mass source parameter and  $M$ is the diffused centralized object. Now, we try to calculate the shape function  as well as the field equations for this particular  modified $f(R)$  gravitational model with respect to its various components.  So, with the help of given   equations (\ref{rho}), (\ref{p}), (\ref{1.3}) and (\ref{1.4}), it is matter of calculation to obtain the shape function as well as other field equations for $(n=\frac{1}{2},1,0)$, specifically. In the case of $n=\frac{1}{2}$ we find $f(R)=-\frac{R}{2}+a R$ which is positive linear term. On the other hand for $n=1$ we have $f(R)=-\frac{R}{2}+a$ which is negative linear term. Finally for $n=0$ we have squared plus negative linear term as $f(R)=-\frac{R}{2}+a R^{2}$.

\subsection{Positive linear model}
For $n=\frac{1}{2}$, we have $f(R)=-\frac{R}{2}+a R$. so the expressions (\ref{rho}), (\ref{1.3}) and (\ref{1.4}) yield
the following shape function,
\begin{equation}\label{1.5}
b(r)=-\frac{M r\sqrt{\phi}}{\pi^{2}(r^{2}+\phi)}+\frac{M\arctan(\frac{r}{\sqrt{\phi}})a}{\pi^{2}}+c,
\end{equation}
where $c$ is an arbitrary integration constant.
 By exploiting equations (\ref{9.0}) and (\ref{1.5}), one can obtain the
 curvature scalar  as follows
\begin{equation}\label{1.6}
R(r)=\frac{2M\sqrt{\phi}((1+a)r^{2}+(-1+a)\phi)}{\pi^{2}r^{2}(r^{2}+\phi)^{2}}.
\end{equation}
Now, the radial and transverse pressure corresponding to  equations (\ref{p})
and (\ref{1.5}) are given by
\begin{eqnarray}
&&P_{r}=\frac{(\frac{1}{2}-a)c-\frac{M\sqrt{\phi}r}{\pi^{2}(\phi+r^{2})}+\frac{M
\arctan(\frac{r}{\sqrt{\phi}})a}{\pi^{2}}}{r^{3}},\label{1.7}\\
&& P_{t}=-\frac{(-1+2a)(-M\sqrt{\phi}r(a\phi+(2+a)r^{2})+(\phi+r^{2})^{2}(a M\arctan(\frac{r}{\sqrt{\phi}})+\pi^{2}c}{4\pi^{2}r^{3}(r^{2}+\phi)^{2}}. \label{1.8}
\end{eqnarray}
Now, these results can be described by plotting graphs.\\

\begin{figure}[h!]
 \begin{center}
 \subfigure[ ]{
 \includegraphics[height=4cm,width=4cm]{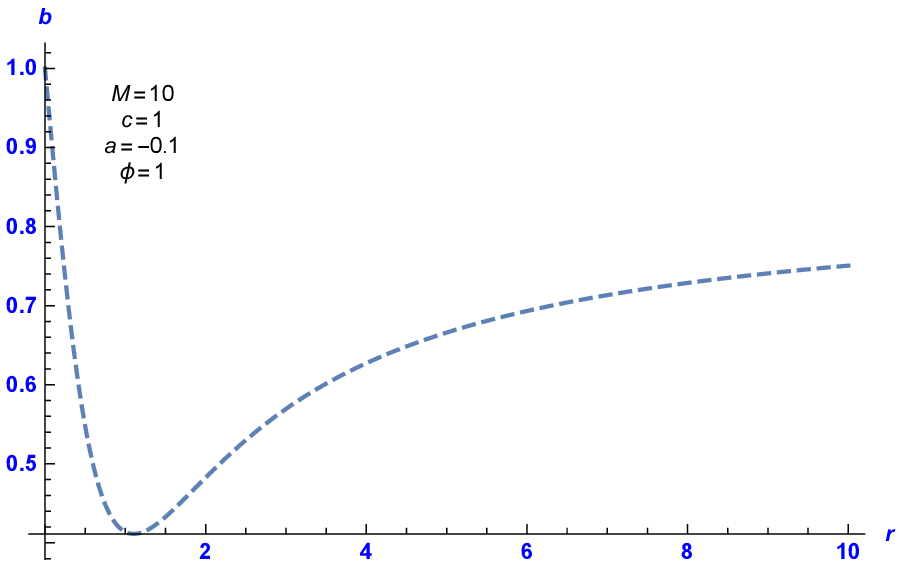}
 \label{1a}}\ \ \ \ \ \ \ \ \ \ \
 \subfigure[ ]{
 \includegraphics[height=4cm,width=4cm]{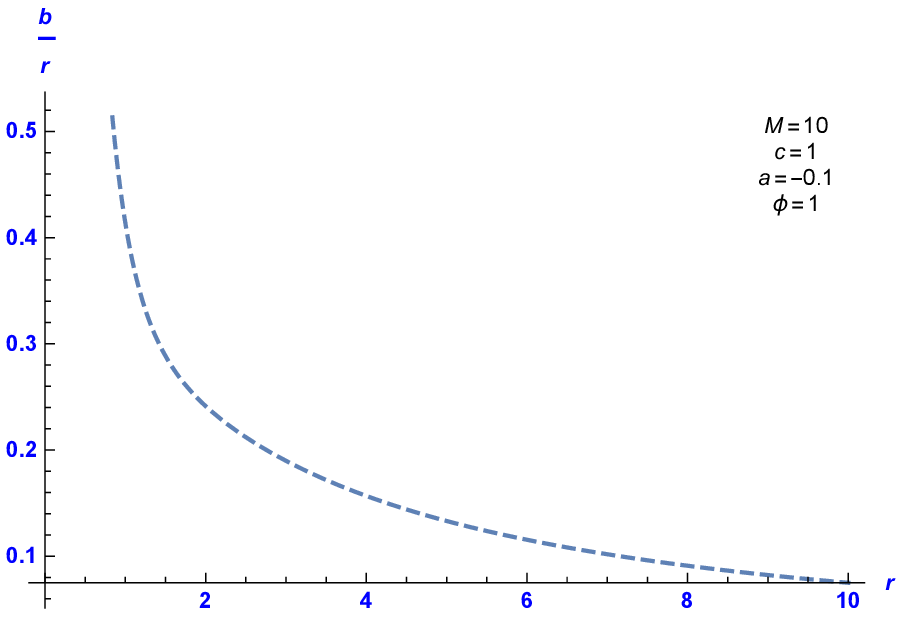}
 \label{1b}}
  \caption{\small{For $n=\frac{1}{2}$. Left plot: shape function  $b$  versus  $r$. Right plot:    $ \frac{b}{r}$ versus  $r$.}}
 \end{center}
 \end{figure}

From the above calculations, it is evident that the considered Lorentz distribution of a particle-like gravitational source is positive for smeared mass source parameter in our calculations. We also
obtain the shape function $b(R)$ for different values of  $n$   with respect to
the modified $f(R)$ gravitational model in relation to wormhole. The redshift
function $\Phi(r)$ is also assumed to be constant at all points.\\

From Fig. \ref{1a}, it is obvious that the shape function increases with respect to $r$. Fig. \ref{1b} reflects that $b/r$ takes asymptotically large value when $r \rightarrow0$, which suggests that the asymptotically
flat condition is being satisfied.\\

\begin{figure}[h!]
 \begin{center}
 \subfigure[]{
 \includegraphics[height=4cm,width=4cm]{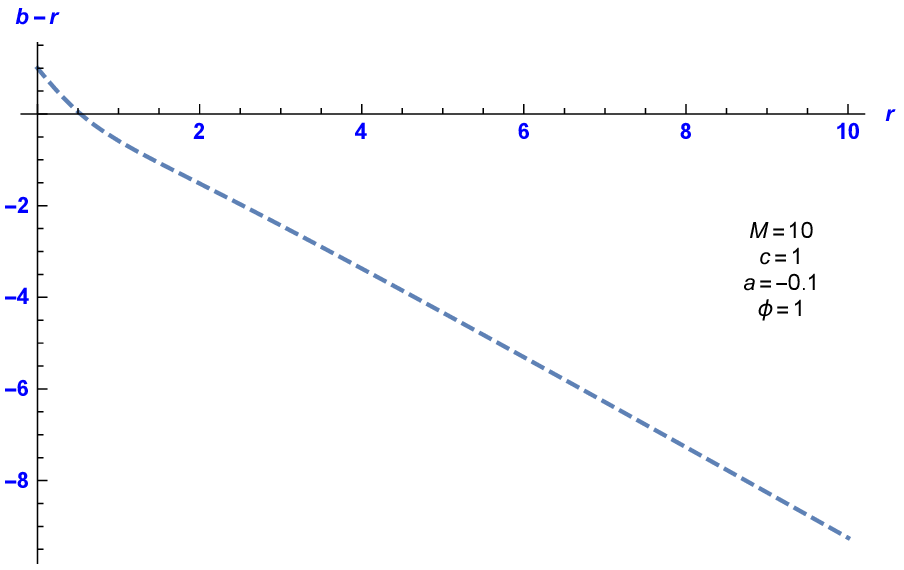}
 \label{2a}}\ \ \ \ \ \ \ \ \ \ \
 \subfigure[]{
 \includegraphics[height=4cm,width=4cm]{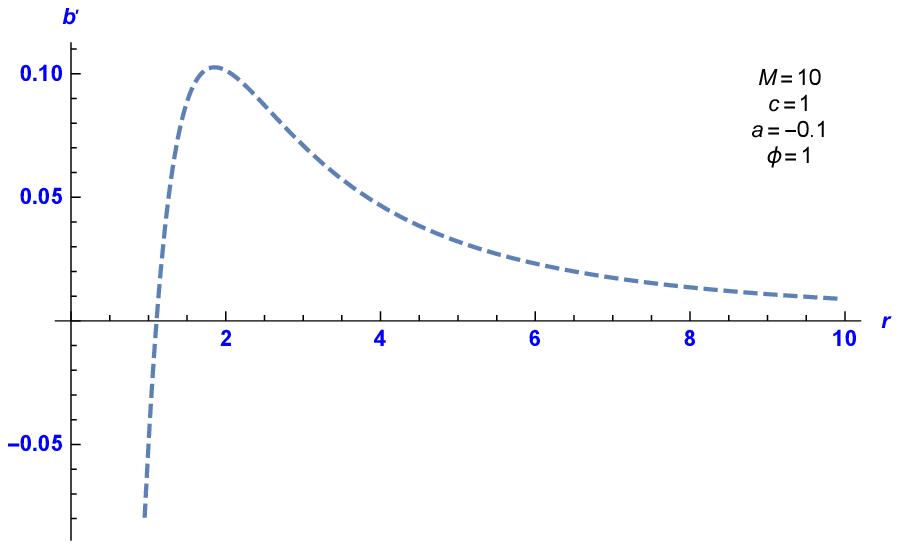}
 \label{2b}}
  \caption{\small{For $n=\frac{1}{2}$. Left plot:  $b-r$   versus $r$  and  right plot:  $b'$
  versus  $r$.
  }}
  \end{center}
 \end{figure}

\begin{figure}[h!]
 \begin{center}
 \subfigure[]{
 \includegraphics[height=4cm,width=4cm]{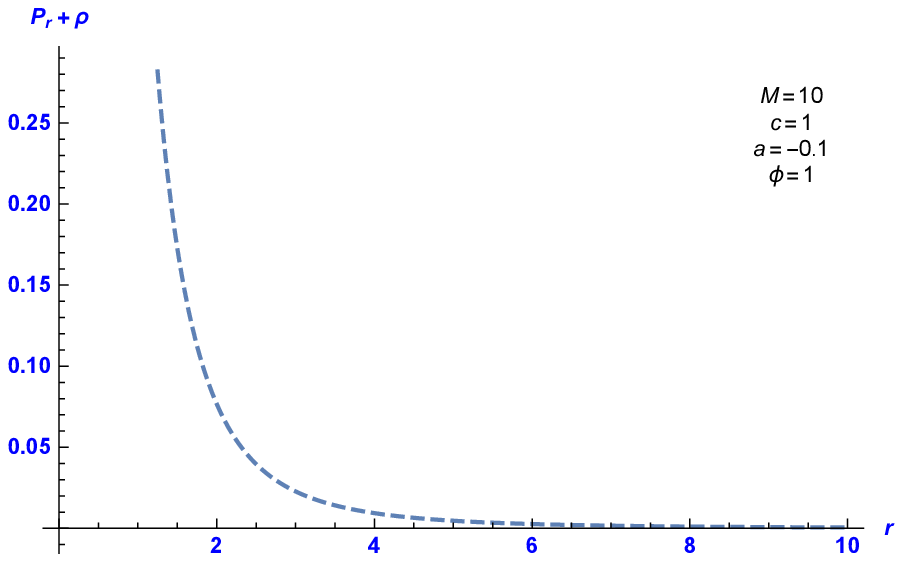}
 \label{3a}}\ \ \ \ \ \ \ \ \ \ \
 \subfigure[]{
 \includegraphics[height=4cm,width=4cm]{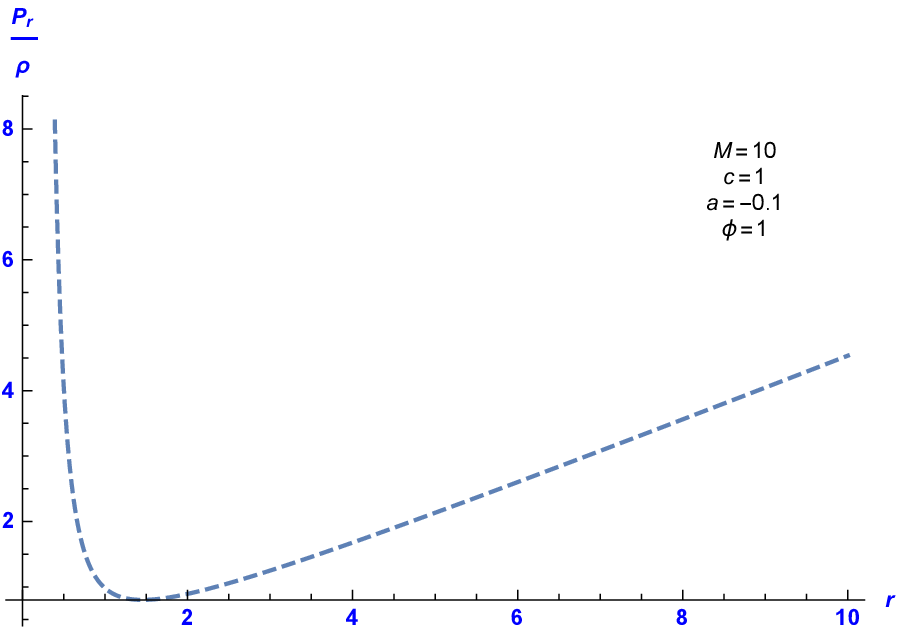}
 \label{3b}}
  \caption{\small{For $n=\frac{1}{2}$. Left plot:   $\rho+P_{r}$ versus $r$. Right plot:   $\frac{P_{r}}{\rho}$ versus $r$.}}
 \end{center}
 \end{figure}

Also, one can see  from the figure  \ref{2a}, the value of throat radius for the wormhole
is estimated as $ (r_0\approx1)$  as curve cuts the $r$ axis at this point.  Fig. \ref{2b} is plotted to check the validity of the condition $b'
(r_0) < 1$. From the plot it can be clearly seen that the  corresponding condition holds.
Consequently, the shape function fulfills all the requirements of warm hole structure. Figs. \ref{3a} and \ref{3b} show $\rho+P_{r}$ and $\frac{P_{r}}{\rho}$ versus $r$.\\

\begin{figure}[h!]
 \begin{center}
 \subfigure[]{
 \includegraphics[height=4cm,width=4cm]{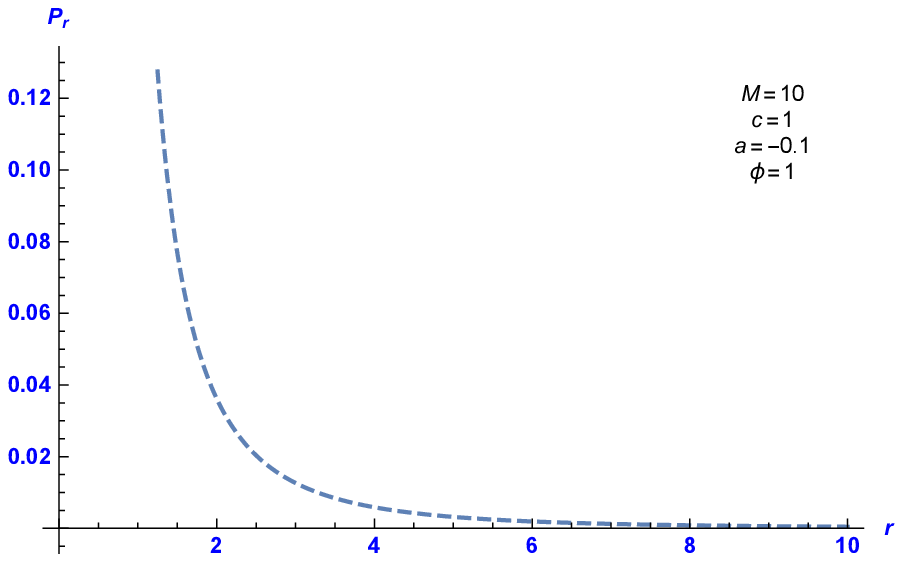}
 \label{4a}}\ \ \ \ \ \ \ \ \ \ \
 \subfigure[]{
 \includegraphics[height=4cm,width=4cm]{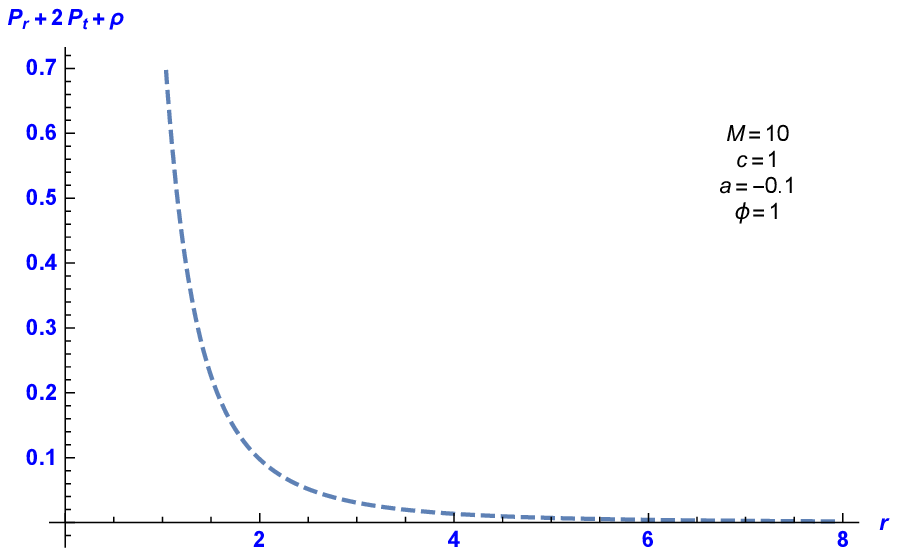}
 \label{4b}}
  \caption{\small{For $n=\frac{1}{2}$. Left plot:  $P_{r}$ versus $r$. Right plot:  and  ($\rho+P_{r}+2P_{t}$) versus $r$. }}
 \end{center}
 \end{figure}

\begin{figure}[h!]
 \begin{center}
 \subfigure[]{
 \includegraphics[height=4cm,width=4cm]{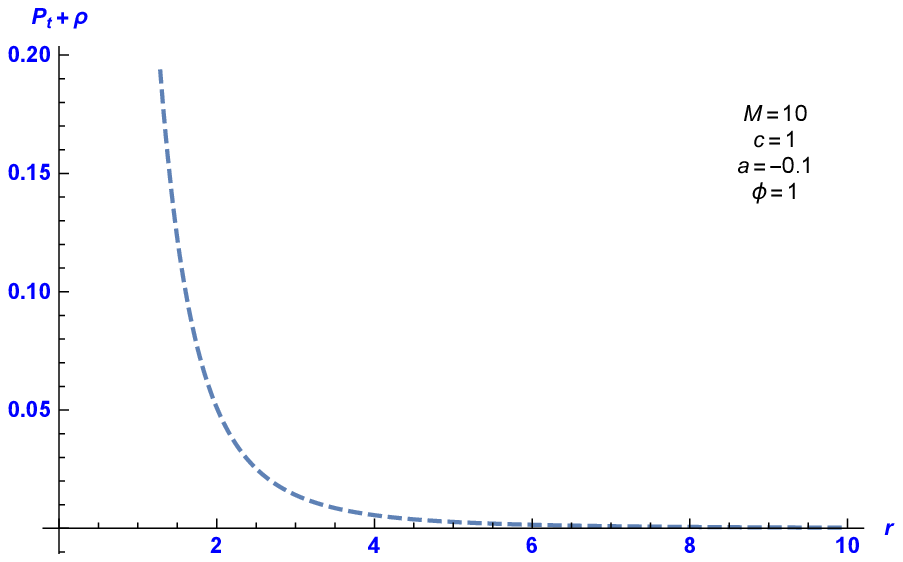}
 \label{5a}}\ \ \ \ \ \ \ \ \ \ \
 \subfigure[]{
 \includegraphics[height=4cm,width=4cm]{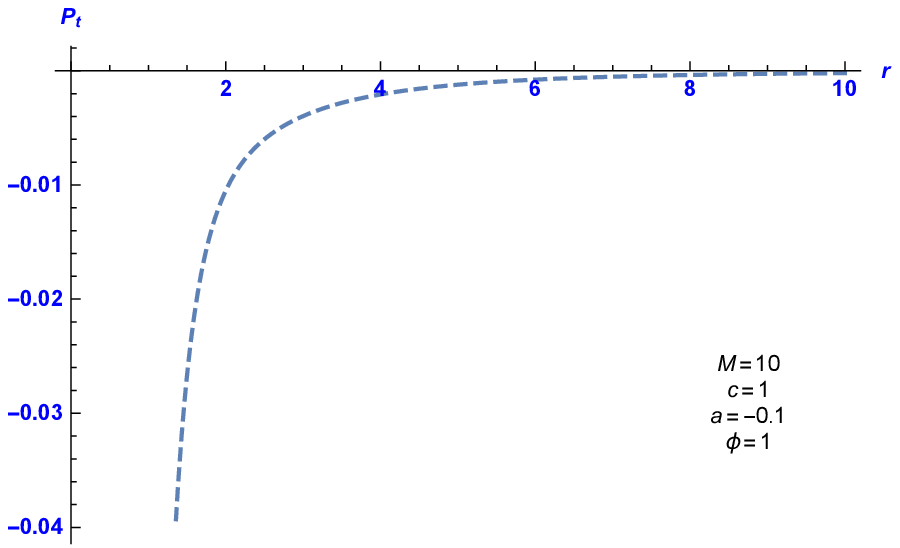}
 \label{5b}}
  \caption{\small{For $n=\frac{1}{2}$. Left plot: $\rho+P_{t}$ versus $r$. Right plot:   $P_{t}$  versus $r$.}}
 \end{center}
 \end{figure}

\begin{figure}[h!]
 \begin{center}
 \subfigure[]{
 \includegraphics[height=4cm,width=4cm]{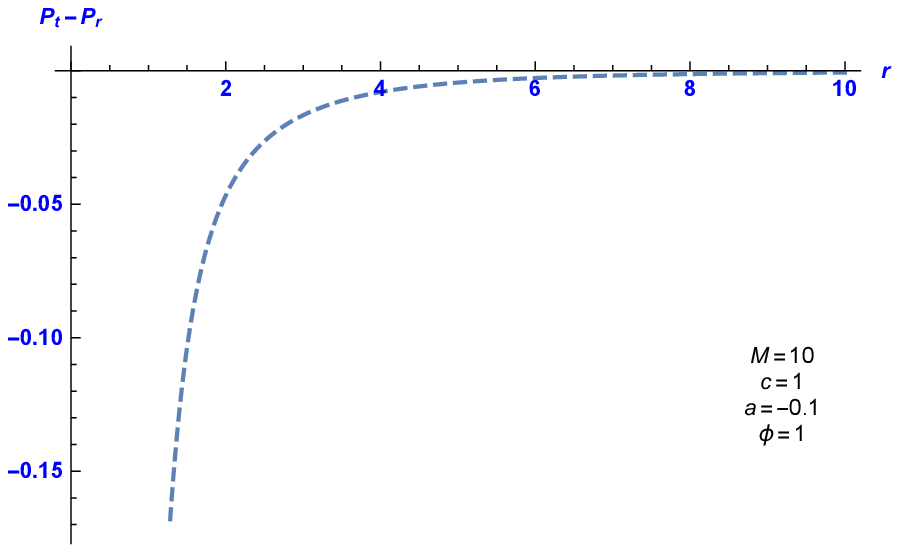}
 \label{6a}}
 \caption{For $n=\frac{1}{2}$. Plot: $P_{t}-P_{r}$ versus $r$.}
 \end{center}
 \end{figure}

The radial pressure and transverse pressure are also plotted in Fig. \ref{4a} and Fig. \ref{5b}, respectively.  We see here that radial pressure  is negative  and transverse pressure is positive. Here, we note that\\
(a) NEC is satisfied if $\rho+P_{r}\geq0$ and $\rho+P_{t}\geq0$;\\
(b) weak energy condition (WEC) is satisfied if $\rho \geq0$, $\rho+P_{r}\geq0$ and $\rho+P_{t}\geq0$;\\
(c)  strong energy condition (SEC) is validated if $\rho+P_{r}\geq0$ , $\rho+P_{t}\geq0$ and $\rho+P_{r}+2P_{t}\geq0$.\\
The equation of state $\omega$ in terms of radial pressure is given by $P_{r}=\omega\rho$ and the anisotropy parameter $\Delta$ is defined by $P_{t}-P_{r}$.
The satisfaction and violation of the energy condition are depicted in  Figs. \ref{3a}, \ref{4b}  and  \ref{5a}. The behavior of equation of state and  anisotropy parameter  can be seen from Figs. \ref{3b} and \ref{6a}, respectively.

\subsection{Negative linear model}
\begin{figure}[h!]
 \begin{center}
 \subfigure[]{
 \includegraphics[height=4cm,width=4cm]{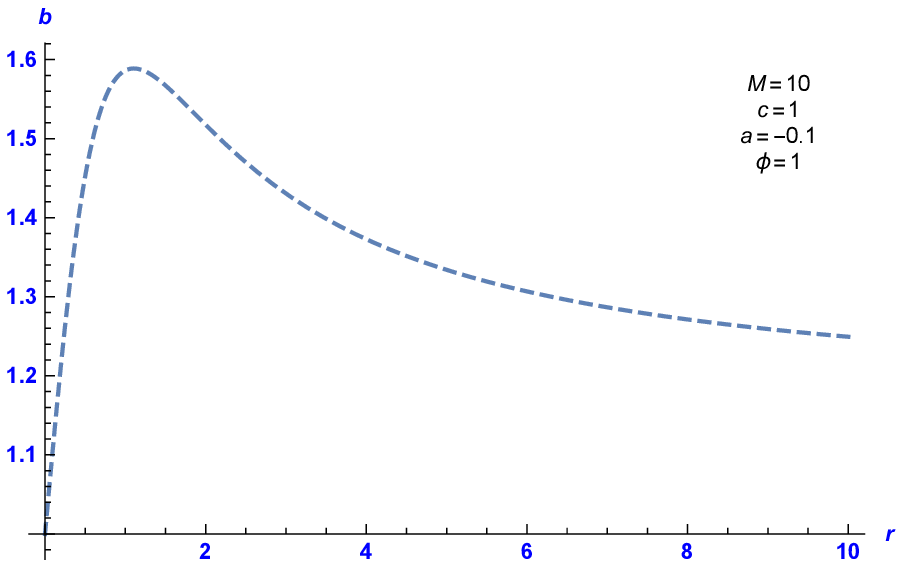}
 \label{7a}}\ \ \ \ \ \ \ \ \ \ \
 \subfigure[]{
 \includegraphics[height=4cm,width=4cm]{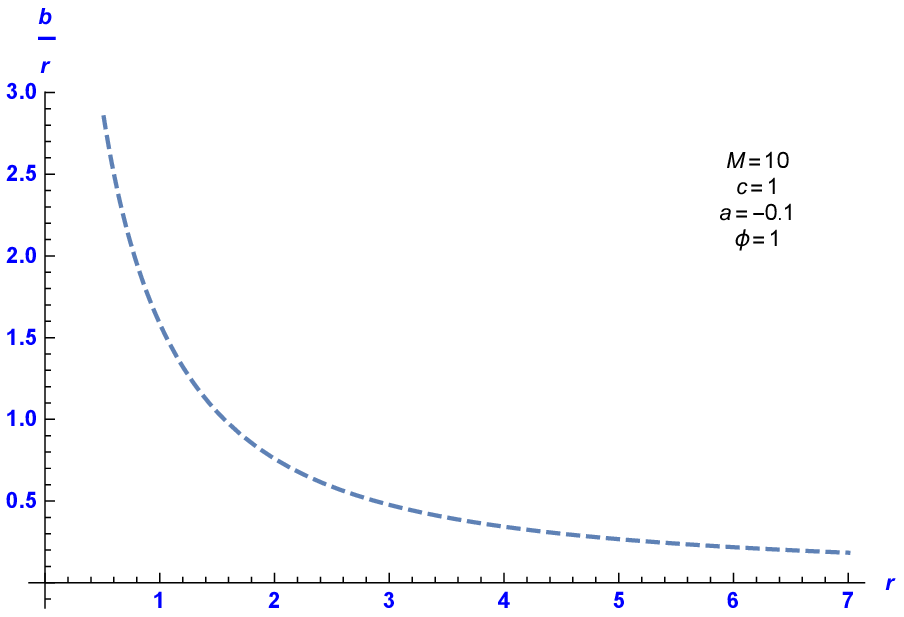}
 \label{7b}}
  \caption{\small{ For $n=1$. Left side: $b$  versus   $r$. Right side:  $\frac{b}{r}$ versus $r$. }}
 \end{center}
 \end{figure}

In the case of $n=1$, we have $f(R)=-\frac{R}{2}+a$ which is a negative linear term. So, we calculate shape function for $n=1$ case as following,
\begin{equation}\label{19}
b(r)=\frac{Mr\sqrt{\phi}}{\pi^{2}(r^{2}+\phi)}-\frac{M\arctan(\frac{r}{\sqrt{\phi}})a}{\pi^{2}}+c,
\end{equation}
where $c$ is an integration constant.
The expression of curvature scalar ($R$) for this case is given by,
\begin{equation}\label{20}
R(r)=-\frac{2M\sqrt{\phi}((1+a)r^{2}+(-1+a)\phi)}{\pi^{2}r^{2}(r^{2}+\phi)^{2}}.
\end{equation}
Exploiting expressions (\ref{p})  and (\ref{19}), the
radial pressure and tangential pressure  are given, respectively, by
\begin{eqnarray}
P_{r}&=&\frac{c+\frac{Mr\sqrt{\phi}}{\pi^{2}(r^{2}+\phi)}-\frac{aM\arctan(\frac{r}{\sqrt{\phi}})}{\pi^{2}}}{2r^{3}},\label{21}\\
P_{t}&=&\frac{c+\frac{Mr\sqrt{\phi}(a\phi+(2+a)r^{2})}{\pi^{2}(r^{2}+\phi)^{2}}-\frac{a M\arctan(\frac{r}{\sqrt{\phi}})}{\pi^{2}}}{4r^{3}}.\label{22}
\end{eqnarray}
After having the required expressions for the case of  $n=1$,
we can analyze the results by plotting graphs.\\
For this case, in contrast to the previous case, we observe that the shape function is a
decreasing function of $r$ as depicted in Fig. \ref{7a}. Also, behavior of $\frac{b}{r}$ illustrated by Fig. \ref{7b}.\\

\begin{figure}[h!]
 \begin{center}
 \subfigure[]{
 \includegraphics[height=4cm,width=4cm]{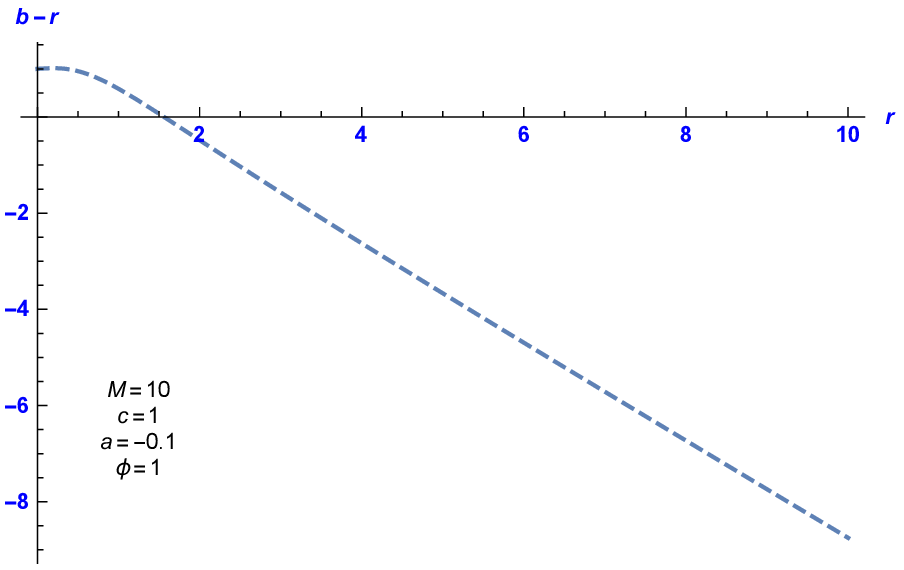}
 \label{8a}}\ \ \ \ \ \ \ \ \ \ \
 \subfigure[]{
 \includegraphics[height=4cm,width=4cm]{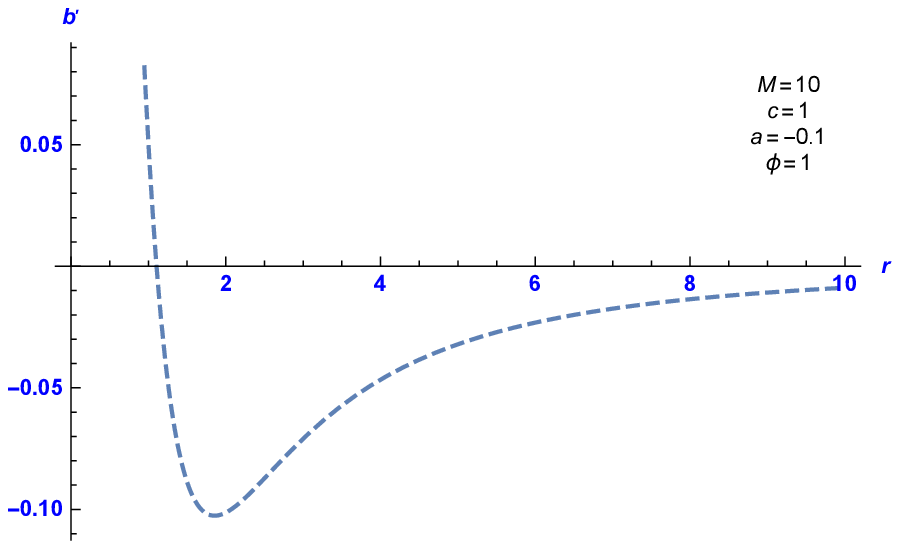}
 \label{8b}}
  \caption{\small{For $n=1$. Left side: $b-r$  versus   $r$. Right side:    $b'$ versus $r$.}}
 \end{center}
 \end{figure}

In Fig. \ref{8a}, we draw $b-r$ versus $r$ to show that is decreasing function of radius. From Fig. \ref{8b}, we see that $b'$
takes only negative values but still follows $b'<1$ condition.\\

\begin{figure}[h!]
 \begin{center}
 \subfigure[]{
 \includegraphics[height=4cm,width=4cm]{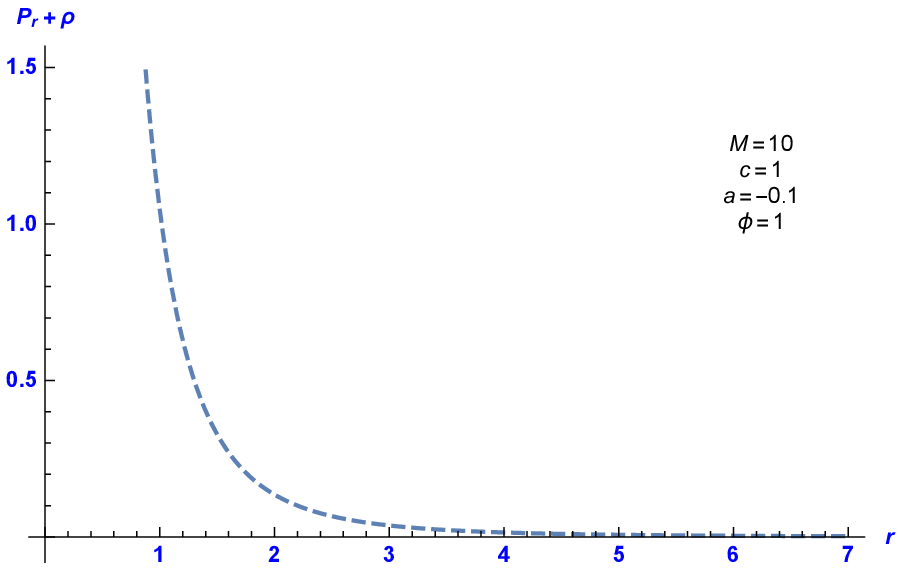}
 \label{9a}}\ \ \ \ \ \ \ \ \ \ \
 \subfigure[]{
 \includegraphics[height=4cm,width=4cm]{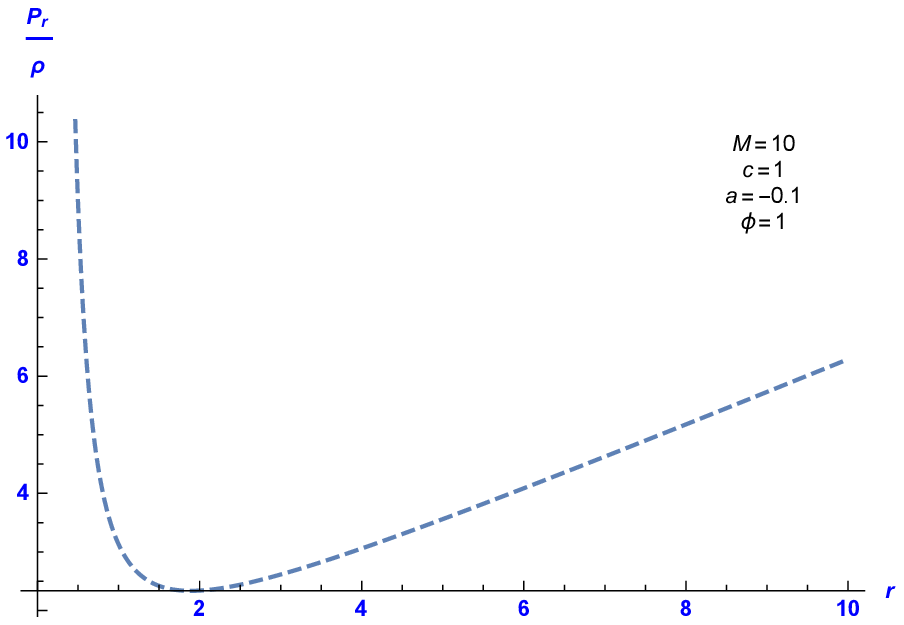}
 \label{9b}}
  \caption{\small{For $n=1$. Left plot:  $\rho+P_{r}$ versus $r$.  Right plot:   $\frac{P_{r}}{\rho}$ versus $r$.  }}
 \end{center}
 \end{figure}

We draw $\rho+P_{r}$ versus $r$ by Fig. \ref{9a} and show that is decreasing function of $r$. From Fig. \ref{9b}, we see that the equation of state shows opposite behavior to that of the previous case.\\

\begin{figure}[h!]
 \begin{center}
 \subfigure[]{
 \includegraphics[height=4cm,width=4cm]{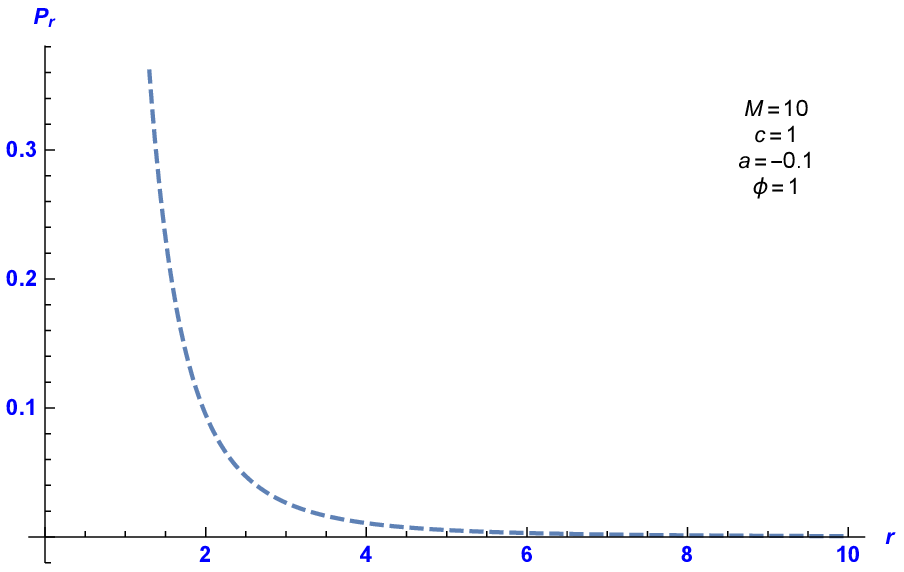}
 \label{10a}}\ \ \ \ \ \ \ \ \ \ \
 \subfigure[]{
 \includegraphics[height=4cm,width=4cm]{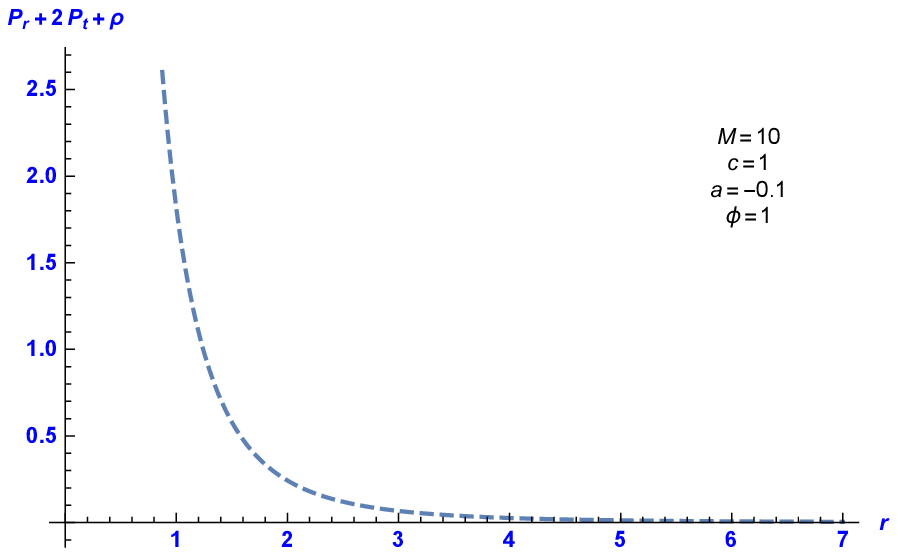}
 \label{10b}}
  \caption{\small {For $n=1$. Left plot: $P_{r}$ versus $r$.  Right plot: $\rho+P_{r}+2P_{t}$ versus $r$. }}
 \end{center}
 \end{figure}

\begin{figure}[h!]
 \begin{center}
 \subfigure[]{
 \includegraphics[height=4cm,width=4cm]{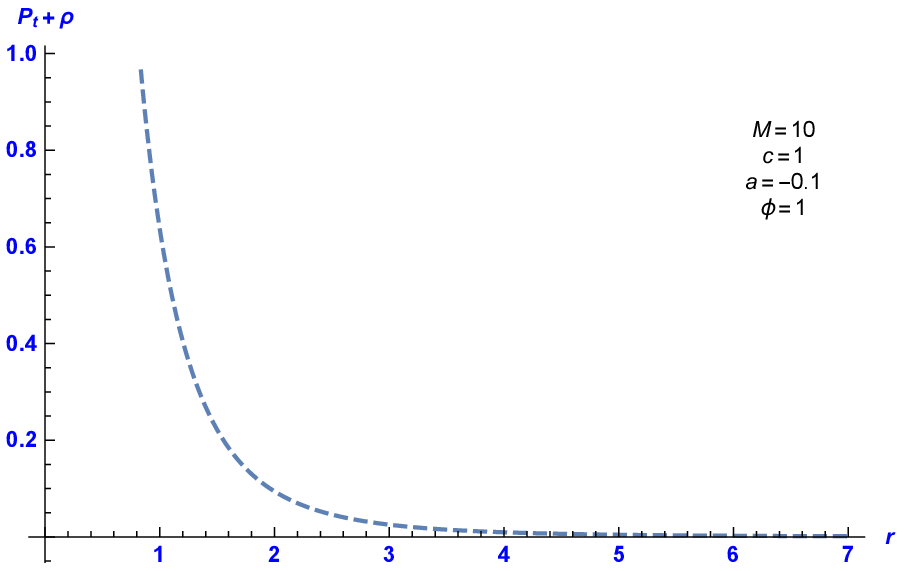}
 \label{11a}}\ \ \ \ \ \ \ \ \ \ \
 \subfigure[]{
 \includegraphics[height=4cm,width=4cm]{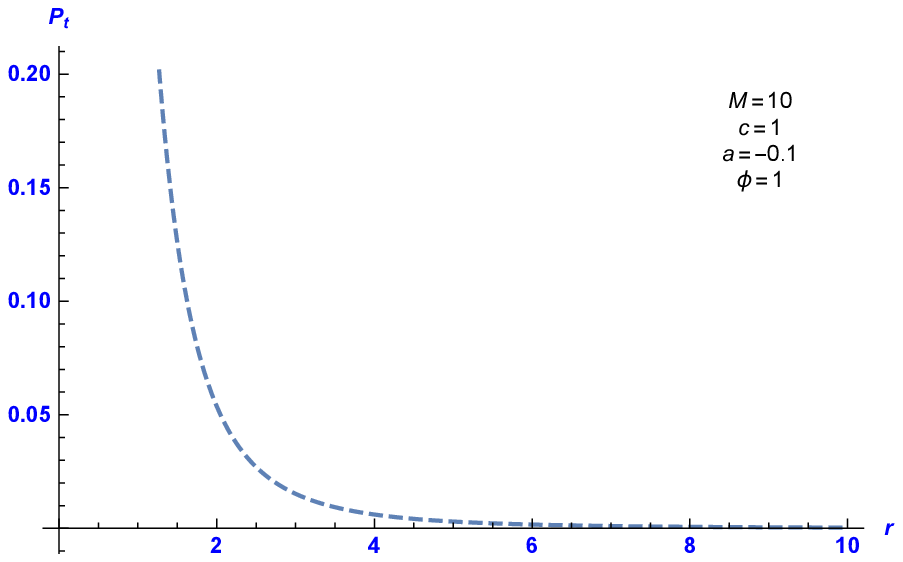}
 \label{11b}}
  \caption{\small{For $n=1$. Left plot: $\rho+P_{t}$ versus $r$.  Right plot:  $P_{t}$ versus $r$.}}
 \end{center}
 \end{figure}

Also, from Figs. \ref{10a} and \ref{11b},
we find that the radial pressure   ($ P_ {r} $)   and the transverse pressure ($ P_ {t} $) are showing opposite behavior as to the case of $n=\frac{1}{2}$.  On the other hand, from Figs. \ref{10b} and \ref{11a}, we see some damping periodic behavior for  $\rho+P_{r}+2P_{t}$ and $\rho+P_{t}$ respectively.\\

\begin{figure}[h!]
 \begin{center}
 \subfigure[]{
 \includegraphics[height=4cm,width=4cm]{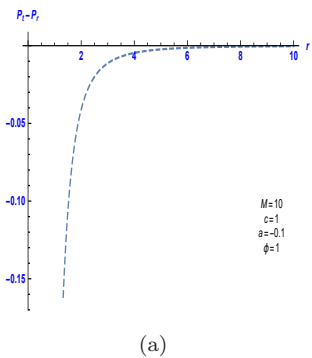}
 \label{12a}}\ \ \ \ \ \ \ \ \ \ \
 \caption{For $n=1$. Plot: $P_{t}-P_{r}$  versus $r$.}
  \end{center}
 \end{figure}

The   NEC, WEC and   SEC  as mentioned in the previous subsection can also be studied in this case also.
 The anisotropy parameter in this case take negative values only as depicted from \ref{12a}.\\

\subsection{Squared plus negative linear model}
In this part, we  investigate another component of the modified $f(R)$ gravitational model, namely $n=0$, which yield $f(R)=-\frac{R}{2}+a R^{2}$.
The computational process of this part is exactly the same as the previous two parts, but we deal with complicated equations which have not analytical solutions. The shape function can be obtained by using the equations (\ref{rho}), (\ref{1.3}) and (\ref{1.4}).  We obtain the values related to curvature scalar $R$, the radial pressure $P_{r}$, tangential pressure $P_{t}$ by using the numerical calculations for the shape function $b(r)$ with respect to equations (\ref{9.0})  and (\ref{p}). The remarkable thing about this modified $f(R)$ gravitational model is that we can obtain two independent solutions for the shape function $b(r)$ corresponding  $n=0$. Here, we discuss  the calculations associated with each value separately. For each case,  the results will be fully described along with  the location of the wormhole throat and the satisfaction/ violation of  the energy conditions.

\subsubsection{The first numerical solutions}
\begin{figure}[h!]
 \begin{center}
 \subfigure[]{
 \includegraphics[height=4cm,width=4cm]{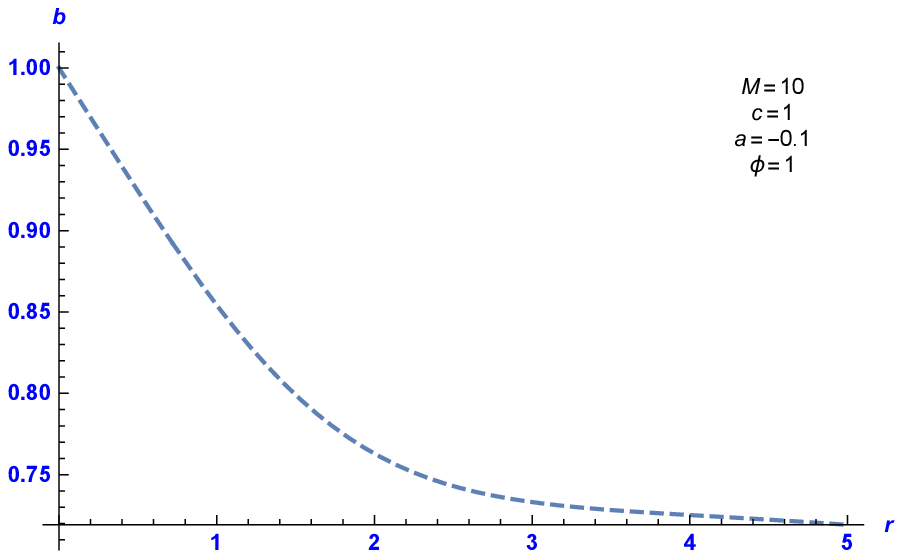}
 \label{13a}}\ \ \ \ \ \ \ \ \ \ \
 \subfigure[]{
 \includegraphics[height=4cm,width=4cm]{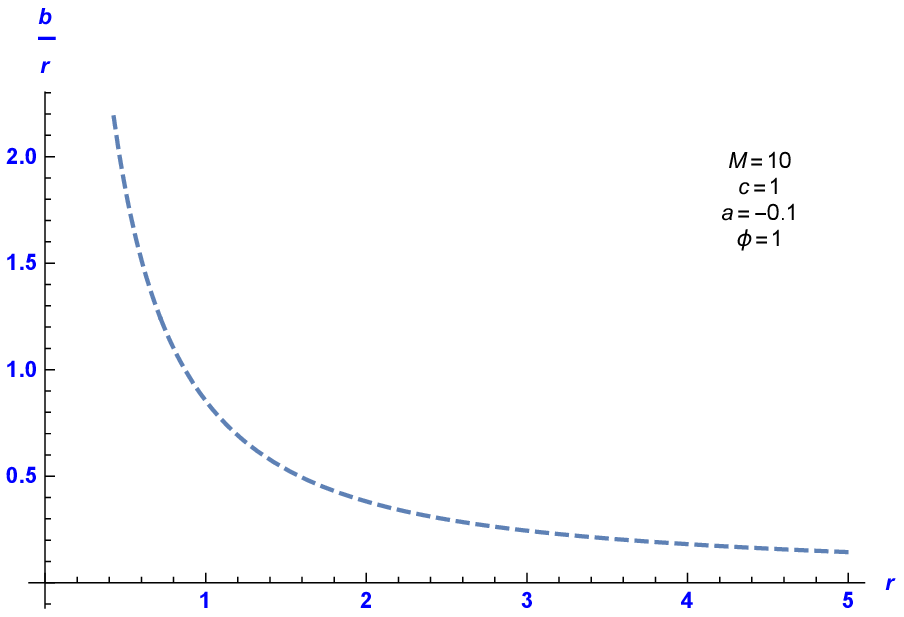}
 \label{13b}}
  \caption{\small{For $n=0$. Left plot: $b$ versus $r$.   Right plot: $\frac{b}{r}$ versus $r$.}}
 \end{center}
 \end{figure}

For this case, we observe that shape function is an increasing function of $r$ as shown in Fig. \ref{13a}. From Fig. \ref{13b}, we find that $b/r$ takes asymptotically large value for small $r$  and asymptotically flat condition is
being satisfied.\\

\begin{figure}[h!]
 \begin{center}
 \subfigure[]{
 \includegraphics[height=4cm,width=4cm]{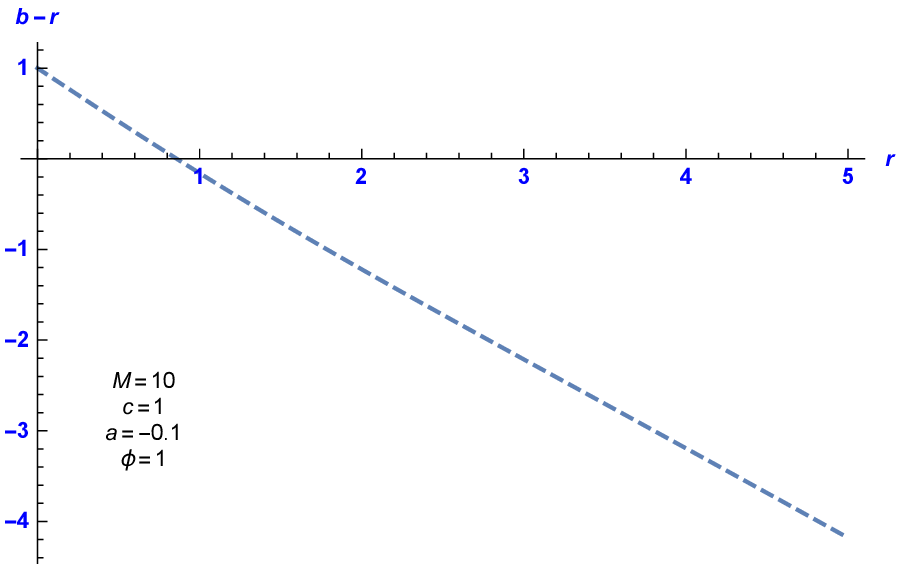}
 \label{14a}}\ \ \ \ \ \ \ \ \ \ \
 \subfigure[]{
 \includegraphics[height=4cm,width=4cm]{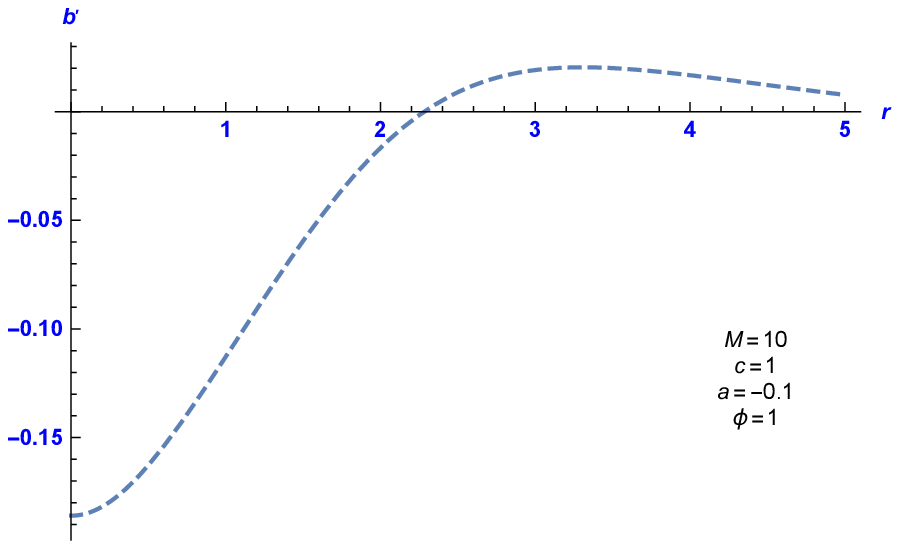}
 \label{14b}}
  \caption{\small{For $n=0$.  Left plot: $b-r$ versus $r$.  Right plot: $b'$ versus $r$.}}
 \end{center}
 \end{figure}

From Fig. \ref{14a}, we see that two throat radii exist for wormhole.
Also, from Fig. \ref{14b}, we see that the shape function does not satisfy the required condition $b'<1$.\\

\begin{figure}[h!]
 \begin{center}
 \subfigure[]{
 \includegraphics[height=4cm,width=4cm]{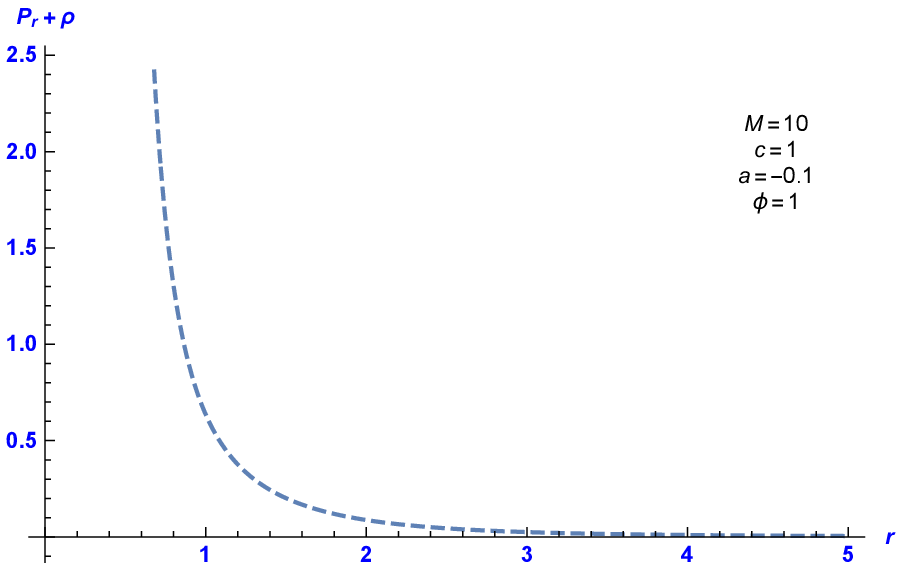}
 \label{15a}}\ \ \ \ \ \ \ \ \ \ \
 \subfigure[]{
 \includegraphics[height=4cm,width=4cm]{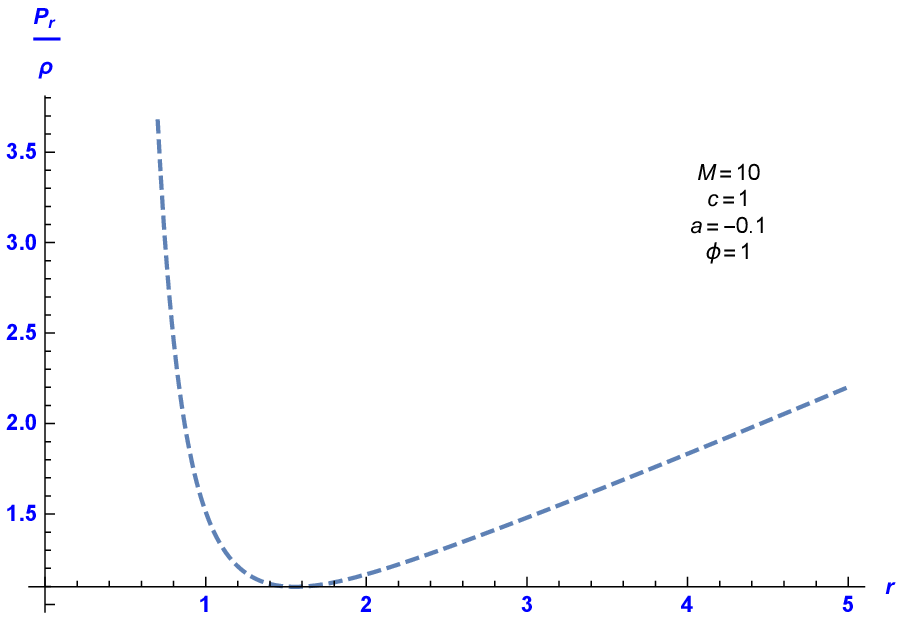}
 \label{15b}}
  \caption{\small{For $n=0$.  Left plot: $\rho+P_{r}$ versus $r$.  Right plot: $\frac{P_{r}}{\rho}$ versus $r$. }}
 \end{center}
 \end{figure}

Fig. \ref{15a} $\rho+P_{r}$ versus $r$ shows that there is a minimum at small radii. From Fig. \ref{15b}, we see that the equation of state shows different behavior to the previous two cases.\\

\begin{figure}[h!]
 \begin{center}
 \subfigure[]{
 \includegraphics[height=4cm,width=4cm]{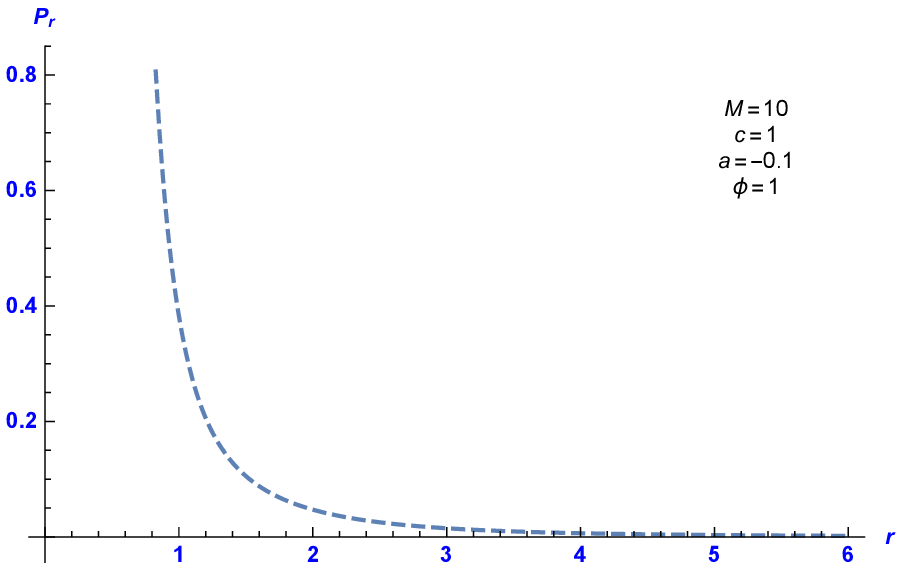}
 \label{16a}}\ \ \ \ \ \ \ \ \ \ \
 \subfigure[]{
 \includegraphics[height=4cm,width=4cm]{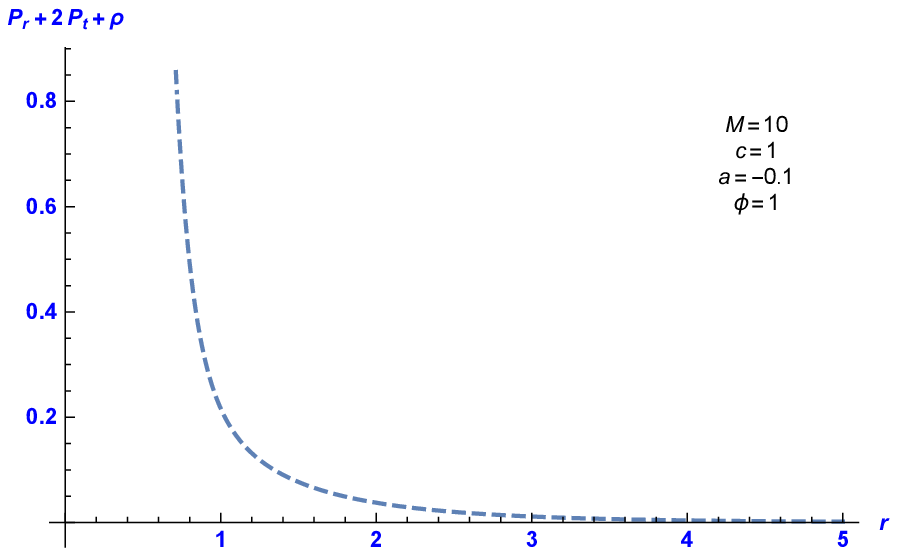}
 \label{16b}}
  \caption{\small{For $n=0$.  Left plot: $P_{r}$ versus $r$.  Right plot: $\rho+P_{r}+2P_{t}$ versus $r$. }}
 \end{center}
 \end{figure}

\begin{figure}[h!]
 \begin{center}
 \subfigure[]{
 \includegraphics[height=4cm,width=4cm]{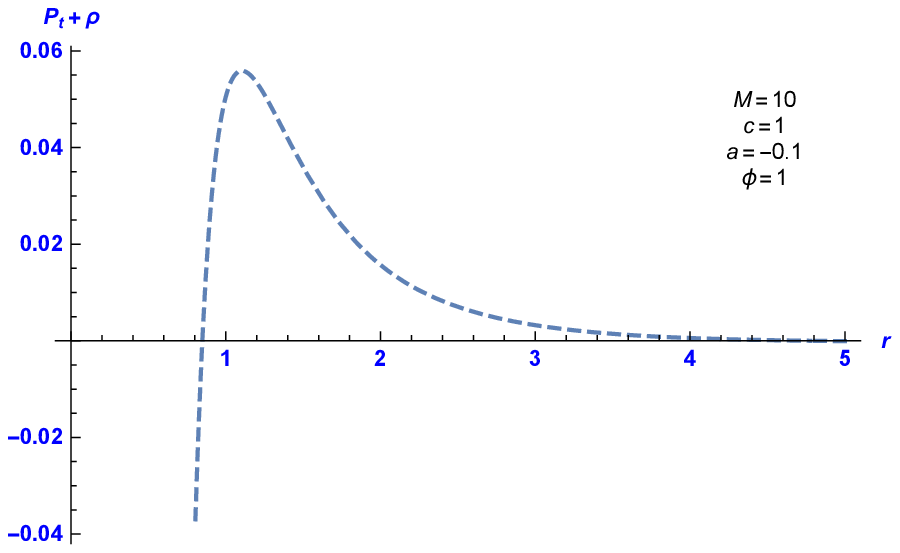}
 \label{17a}}\ \ \ \ \ \ \ \ \ \ \
 \subfigure[]{
 \includegraphics[height=4cm,width=4cm]{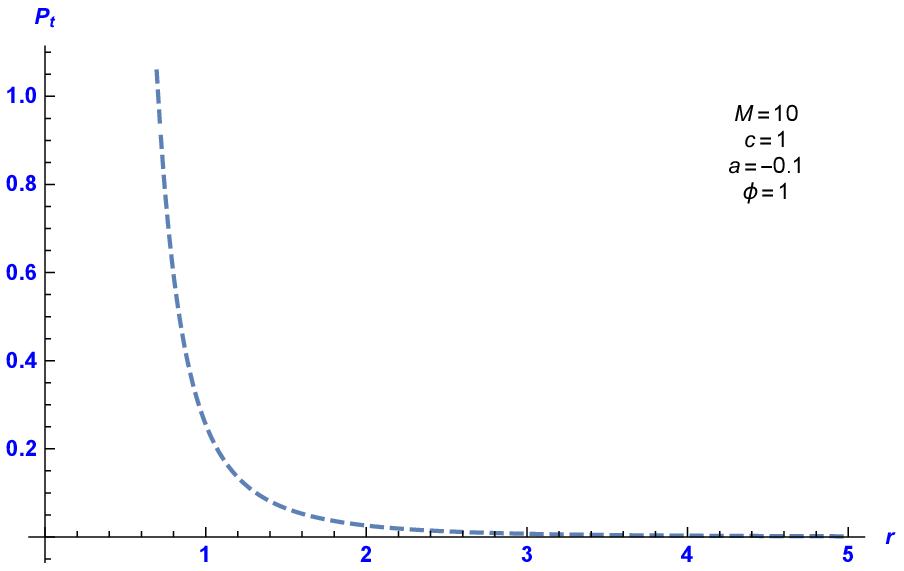}
 \label{17b}}
  \caption{\small{For $n=0$. Left plot: $\rho+P_{t}$ versus $r$.  Right plot: $P_{t}$ versus $r$.}}
 \end{center}
 \end{figure}

In the plots  \ref{16a}  and  \ref{17b}, we can see that the radial pressure  ($ P_ {r} $) and  the transverse pressure ($ P_ {t} $) are showing opposite behavior. The region where  NEC, WEC  and SEC are satisfied/violated can be studied from the plots \ref{15a}, \ref{16b} and \ref{17a}.
 The $\rho+P_r$ and $\rho+P_r+2P_t$ showing similar behavior in contrast to above cases.\\

\begin{figure}[h!]
 \begin{center}
 \subfigure[]{
 \includegraphics[height=4cm,width=4cm]{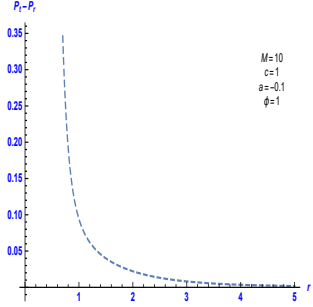}
 \label{18a}}\ \ \ \ \ \ \ \ \ \ \
 \caption{For $n=0$. Plot: $P_{t}-P_{r}$ versus $r$.}
 \end{center}
 \end{figure}

Anisotropy parameter curve shows behavior like  nuclear potential as depicted in Fig. \ref{18a}, which including a minimum at small radii.

\subsubsection{The second numerical solutions}
\begin{figure}[h!]
 \begin{center}
 \subfigure[]{
 \includegraphics[height=4cm,width=4cm]{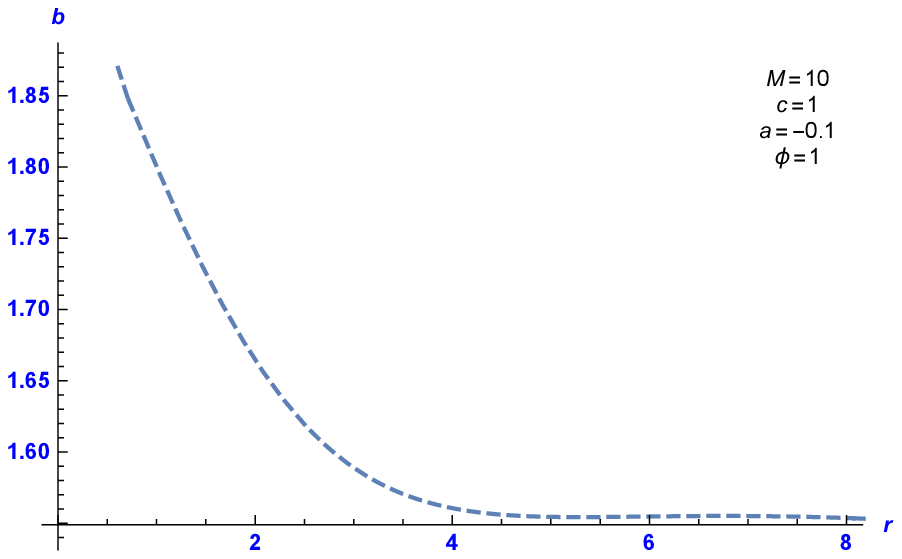}
 \label{19a}}\ \ \ \ \ \ \ \ \ \ \
 \subfigure[]{
 \includegraphics[height=4cm,width=4cm]{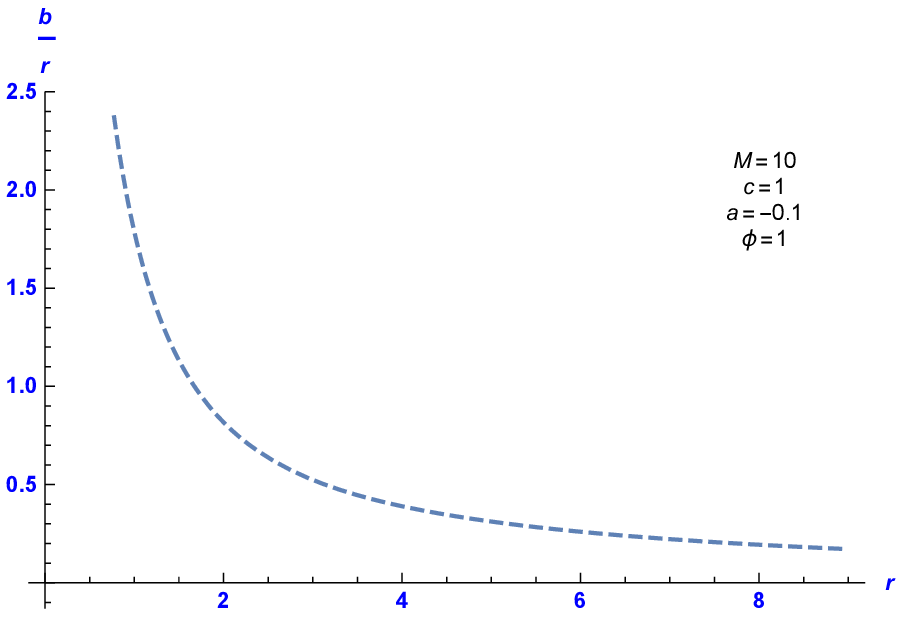}
 \label{19b}}
  \caption{\small{For $n=0$. Left plot: $b$ versus $r$.  Right plot: $\frac{b}{r}$ versus $r$. }}
 \end{center}
 \end{figure}

In the second case of $n=0$, on the contrary,  we observe that shape function is a decreasing function of $r$  as shown in Fig. \ref{19a}. From Fig. \ref{19b}, we find that $b/r$ takes asymptotically large value for small $r$  and asymptotically flat condition is
being satisfied here also.\\

\begin{figure}[h!]
 \begin{center}
 \subfigure[]{
 \includegraphics[height=4cm,width=4cm]{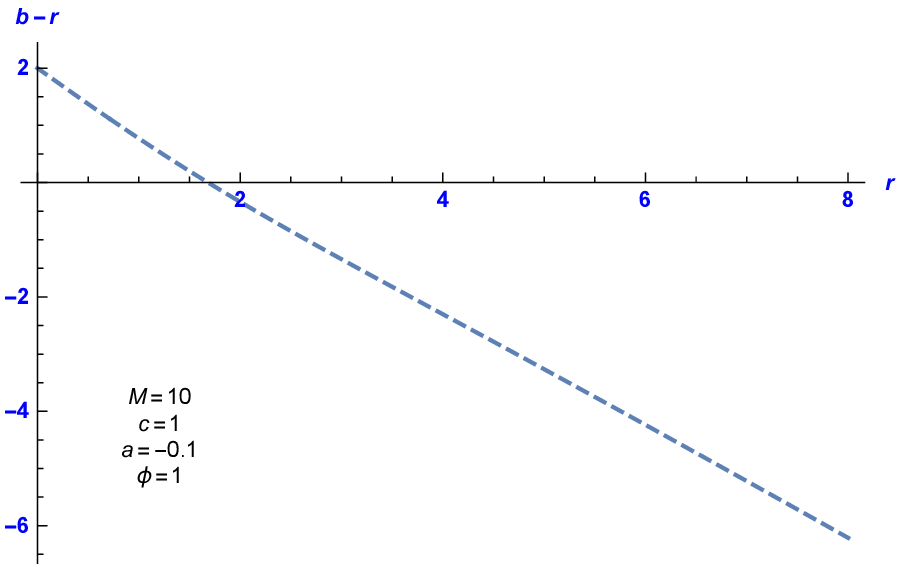}
 \label{20a}}\ \ \ \ \ \ \ \ \ \ \
 \subfigure[]{
 \includegraphics[height=4cm,width=4cm]{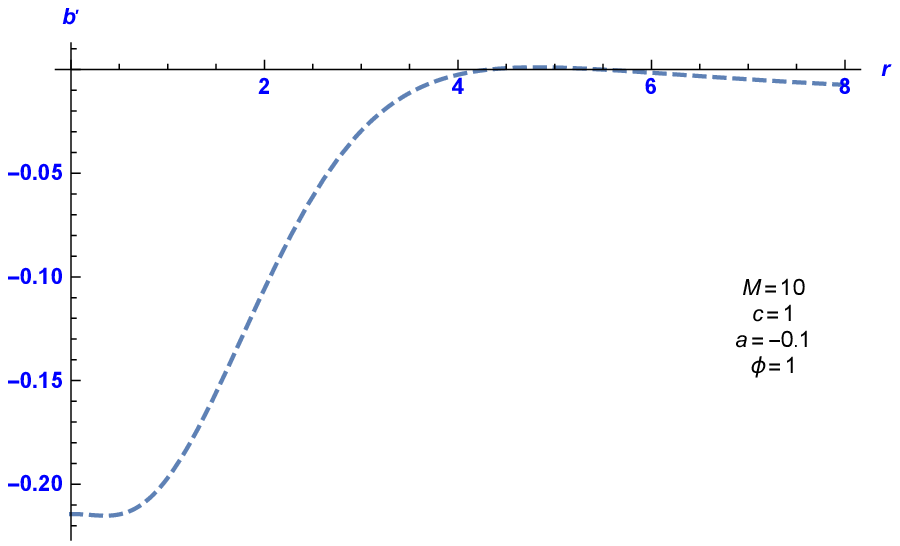}
 \label{20b}}
  \caption{\small{For $n=0$. Left plot: $b-r$ versus $r$.  Right plot: $b'$ versus $r$. }}
 \end{center}
 \end{figure}

From Fig. \ref{20a}, we see that the throat radius take value $r\approx1$ for wormhole.
Moreover, from Fig. \ref{20b}, we see that the shape function   satisfies the required condition $b'<1$.\\

\begin{figure}[h!]
 \begin{center}
 \subfigure[]{
 \includegraphics[height=4cm,width=4cm]{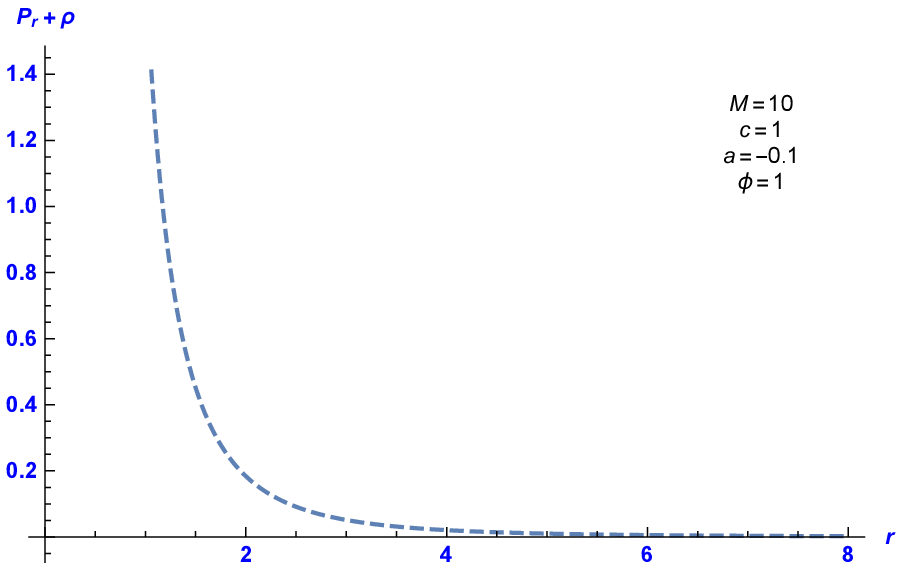}
 \label{21a}}
 \subfigure[]{
 \includegraphics[height=4cm,width=4cm]{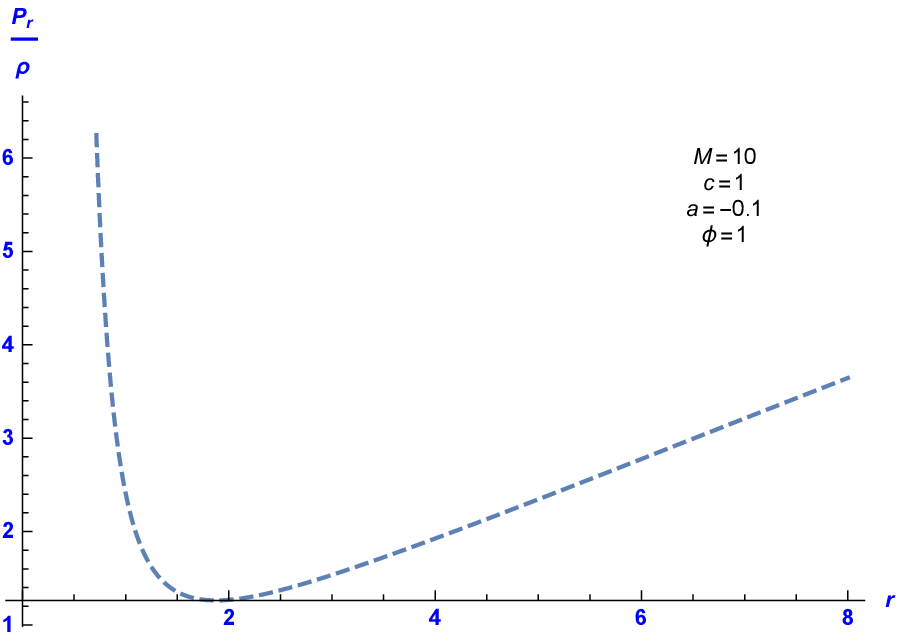}
 \label{21b}}
  \caption{\small{For $n=0$. Left plot: $\rho+P_{r}$ versus $r$.  Right plot: $\frac{P_{r}}{\rho}$ versus $r$. }}
 \end{center}
 \end{figure}

\begin{figure}[h!]
 \begin{center}
 \subfigure[]{
 \includegraphics[height=4cm,width=4cm]{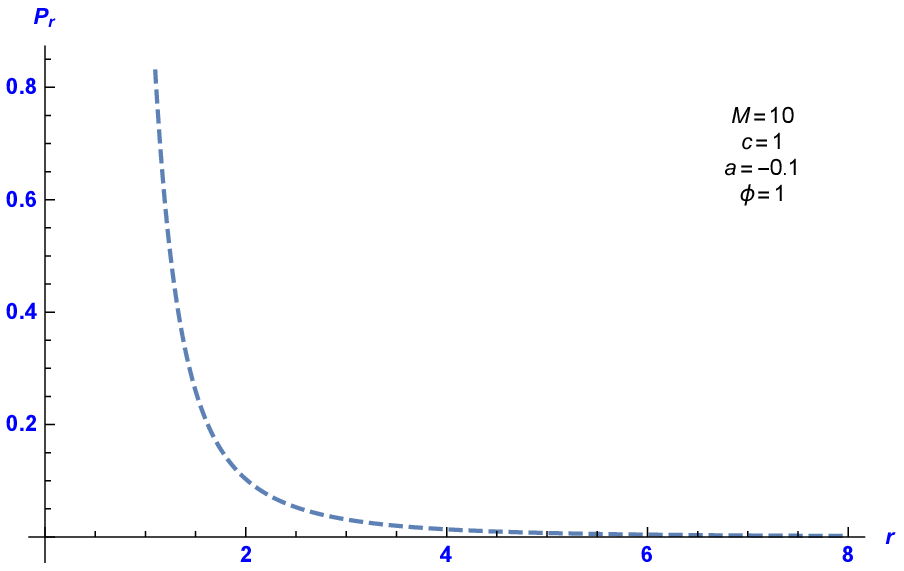}
 \label{22a}}\ \ \ \ \ \ \ \ \ \ \
 \subfigure[]{
 \includegraphics[height=4cm,width=4cm]{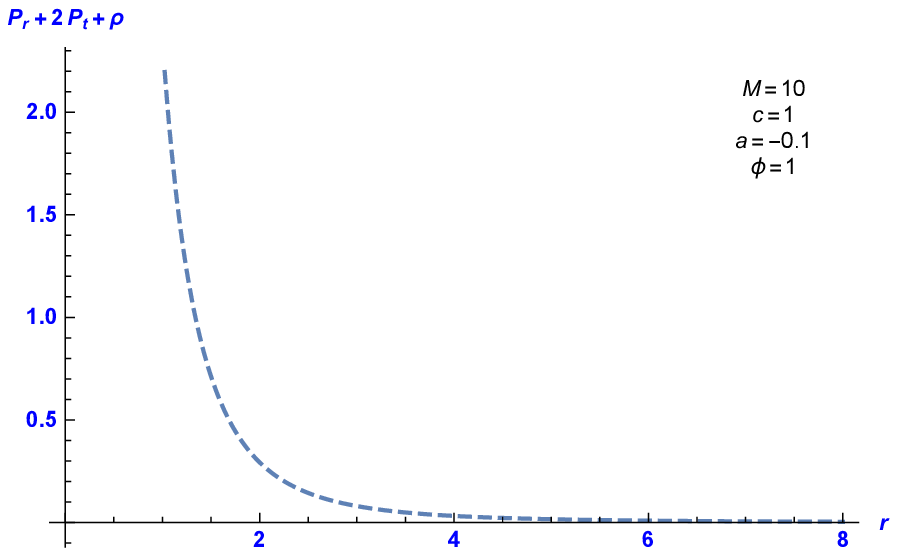}
 \label{22b}}
  \caption{\small{For $n=0$. Left plot: $P_{r}$ versus $r$.  Right plot: $\rho+P_{r}+2P_{t}$ versus $r$. }}
 \end{center}
 \end{figure}

\begin{figure}[h!]
 \begin{center}
 \subfigure[]{
 \includegraphics[height=4cm,width=4cm]{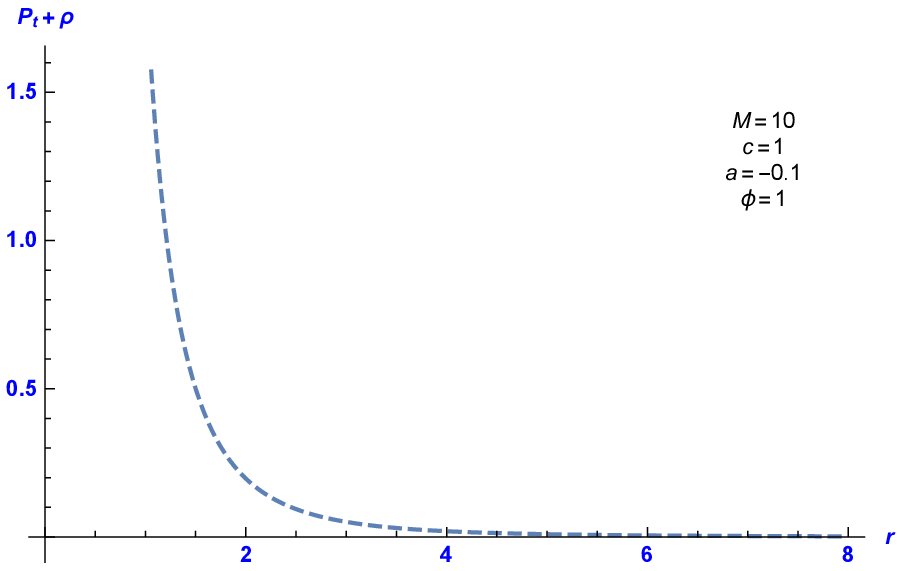}
 \label{23a}}\ \ \ \ \ \ \ \ \ \ \
 \subfigure[]{
 \includegraphics[height=4cm,width=4cm]{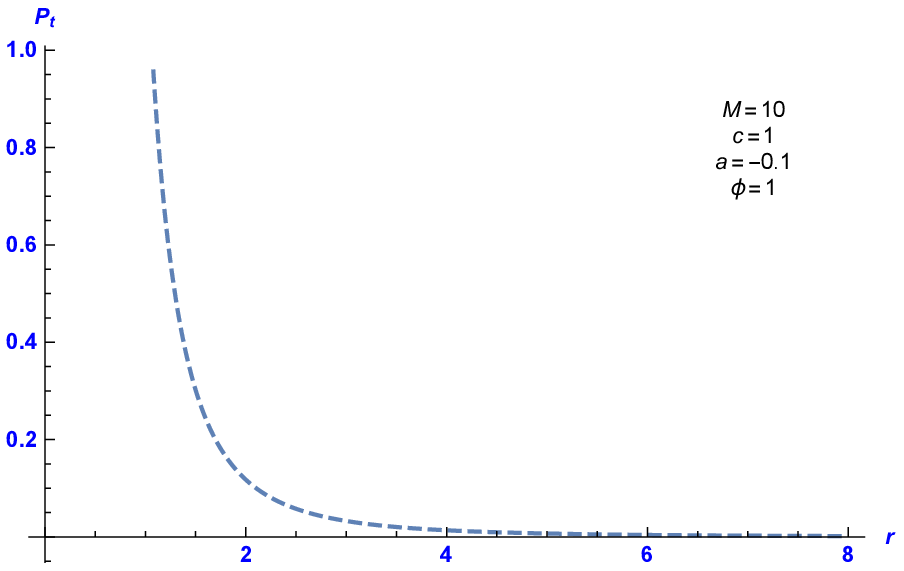}
 \label{23b}}
  \caption{\small{For $n=0$. Left plot: $\rho+P_{t}$ versus $r$.  Right plot: $P_{t}$ versus $r$. }}
 \end{center}
 \end{figure}

\begin{figure}[h!]
 \begin{center}
 \subfigure[]{
 \includegraphics[height=4cm,width=4cm]{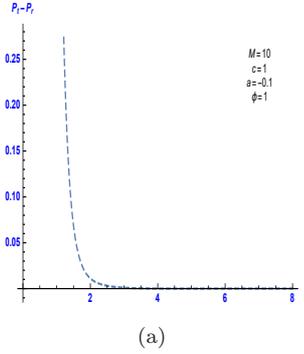}
 \label{24a}}
 \caption{For $n=0$. Plot: $P_{t}-P_{r}$ versus $r$.}
 \end{center}
 \end{figure}

From \ref{21b}, we see that the equation of state takes asymptotically small value for small $r$. In the plots  \ref{22a}  and  \ref{23b}, we can see that the radial pressure  ($ P_ {r} $) and  the transverse pressure ($ P_ {t} $) show opposite nature to the first case of $n=0$. Anisotropy parameter curve shows similar behavior of equation of state as depicted in figure \ref{24a}. The region where  NEC, WEC  and SEC are satisfied/violated can be studied from the plots \ref{21a}, \ref{22b} and \ref{23a}. The $\rho+P_r$ and $\rho+P_r+2P_t$ showing similar behavior in contrast to $n=1/2,1$ cases.\\

We end up this section by stating that  these concepts expressed here can be applied to other models as well. The more interesting point here is that  by introducing different wormholes and recently proposed conditions  such as weak gravity conjecture,  one can shed new light on the structure associated with wormholes.\\

At the end of this section, the energy conditions for different $n$ for this model can be briefly expressed in the following form.\\
$\bullet$ for $n=\frac{1}{2}$, $NEC$, $WEC$, and $SEC$ are satisfied.\\
$\bullet$ for $n=1$, $NEC$, $WEC$, and $SEC$ are satisfied.\\
$\bullet$ for $n=0(I)$ , $NEC$, $WEC$, and $SEC$ are satisfied in $(r>1)$\\
$\bullet$ for $n=0(II)$ , $NEC$, $WEC$, and $SEC$ are satisfied

\section{Specific solutions  for  the power law model}\label{s3}
In this section, we consider another model where $f(R)$ specifies as\cite{38,39,40,41}
\begin{equation}\label{23}
f(R)=\frac{\alpha}{R^{n}},
\end{equation}
where $\alpha$ and $n$ are some constants. In fact,  our main goal is to obtain the field equations  using non-commutative geometry with Lorentzian distribution. So, we express the energy density $(\rho)$ associated with spherical symmetry and particle-like gravitational source Which is mentioned in equation (\ref{1.4})\cite{35}: Now, with respect to equations (\ref{9.0}) and (\ref{23}) we have
\begin{equation}\label{25}
f(R)=\frac{\alpha}{\left(2\frac{b'}{r^{2}}\right)^{n}}.
\end{equation}
Now by plugging  expression (\ref{1.4}) and (\ref{25}) in equation (\ref{rho}), we obtain the shape function $b(r)$  given  by
\begin{equation}\label{26}
b(r)=\int \pi^{\frac{2}{n-1}}\left(\frac{2^{n}M(\frac{1}{r^{2}})^{n-1}\sqrt{\phi}}{\alpha(r^{2}+\phi)^{2}}\right)^{\frac{1}{n-1}}dr.
\end{equation}
Then, we shall study the specific cases corresponding to different values of $n$.
\subsection{Shape function, energy condition and equation of states for $ n=\frac{1}{2}$}
For $n=\frac{1}{2}$, the shape function  (\ref{26})  simplifies to
\begin{equation}\label{27}
b(r)=\frac{M^{2}}{24\alpha^{2}\pi^{4}\phi^{\frac{3}{2}}}\left[-\frac{\sqrt{\phi}r(\phi-3r^{2})(3\phi+r^{2})}{(\phi+r^{2})^{3}}+3\arctan\left(\frac{r}{\sqrt{\phi}}\right)\right]+c,
\end{equation}
where $c$ is an integration constant.

Utilizing expressions (\ref{27}) and  (\ref{9.0}), we obtain the specific  form of curvature scalar
as following:
\begin{equation}\label{28}
R(r)=\frac{4M^{2}\phi}{\alpha^{2}\pi^{4}(\phi+r^{2}){4}}
\end{equation}
With the help of expressions (\ref{27}) and (\ref{p}), the radial pressure and tangential pressure are computed, respectively, as
\begin{eqnarray}\label{29}
P_{r}&=&\frac{A}{B},\\
A&=&-\alpha\sqrt{\frac{M^{2}\phi^{2}}{\alpha^{2}(\phi+r^{2})^{4}}}
\left(-\sqrt{\phi}(24\alpha^{2}c\pi^{4}\phi(\phi+r^{2})^{4}+M^{2}r(-\phi+r^{2})(3\phi^{2}+14\phi r^{2}+3r^{4})\right)\nonumber\\
&-&3M^{2}(\phi+r^{2})^{4}\arctan(\frac{r}{\sqrt{\phi}}),\\
B&=&32M^{2}\pi^{2}\phi^{\frac{5}{2}}r^{3},
\end{eqnarray}
\begin{eqnarray}\label{30}P_{t}&=&\frac{C}{D},\\
C&=&-\alpha\sqrt{\frac{M^{2}\phi}{\alpha^{2}(\phi+r^{2})^{4}}}(-\sqrt{\phi}r(16\alpha^{2}\pi^{4}\phi)(\phi+r^{2})^{4}+M^{2}(\phi-r^{2})\nonumber\\
&&\times(3\phi^{2}+14\phi r^{2}+3r^{4})-3M^{2}(\phi+r^{2})^{4}\arctan(\frac{r}{\sqrt{\phi}})),\\
D&=&32\pi^{2}M^{2}\phi^{\frac{5}{2}}r^{3}.
\end{eqnarray}
Now, to analyze  these results, we  plot graphs.\\

\begin{figure}[h!]
 \begin{center}
 \subfigure[]{
 \includegraphics[height=4cm,width=4cm]{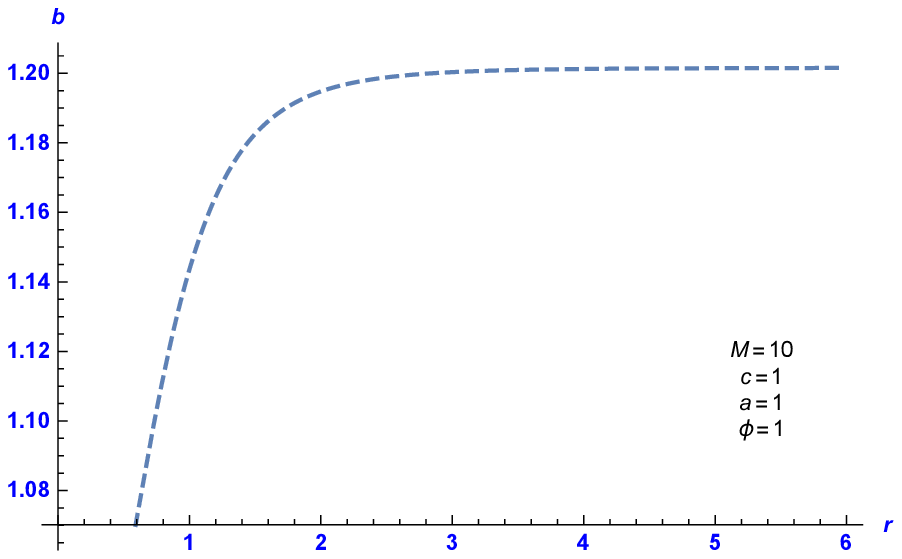}
 \label{26a}}\ \ \ \ \ \ \ \ \ \ \
 \subfigure[]{
 \includegraphics[height=4cm,width=4cm]{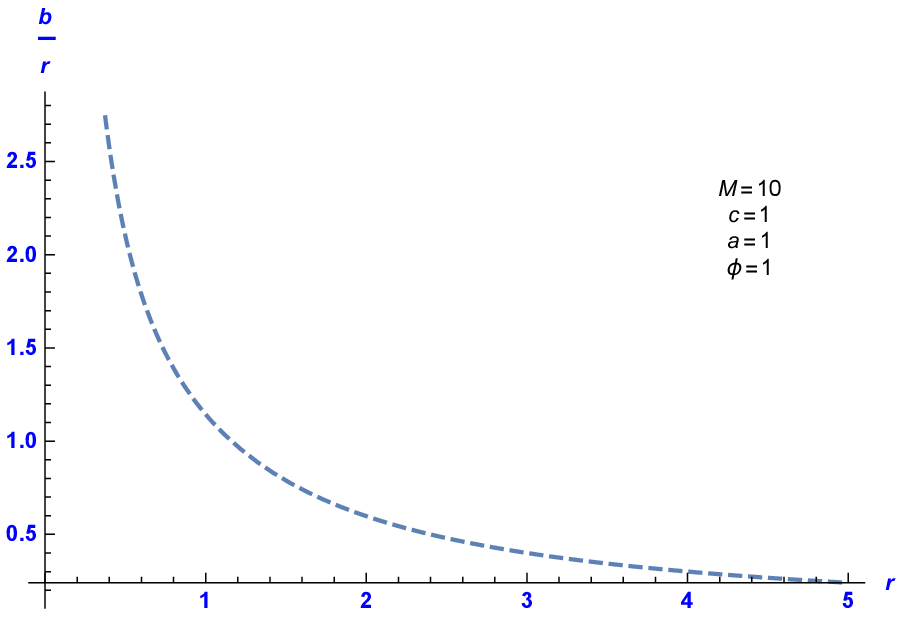}
 \label{26b}}
  \caption{\small{For $ n=\frac{1}{2}$. Left plot: $b$ versus $r$.  Right plot: $\frac{b}{r}$ versus $r$.  }}
 \end{center}
 \end{figure}

From Fig. \ref{25a}, it is obvious that the shape function increases rapidly and saturates after a point along with $r$. In Fig. \ref{25b}, we see that similar to other cases here also $b/r$ takes asymptotically large value when $r \rightarrow0$, which suggests that the asymptotically
flat condition   is being satisfied.\\
In this case of $f(R)$ model,   the value of throat radius for the wormhole in this model is $ (r_0\approx 1.2)$ as depicted in the figure  \ref{26a}.
In  Fig. \ref{26b}, justifies the validity of the condition $b'(r_0) < 1$ and, hence,  the shape function satisfies all the requirements of warm hole structure.\\

 \begin{figure}[h!]
 \begin{center}
 \subfigure[]{
 \includegraphics[height=4cm,width=4cm]{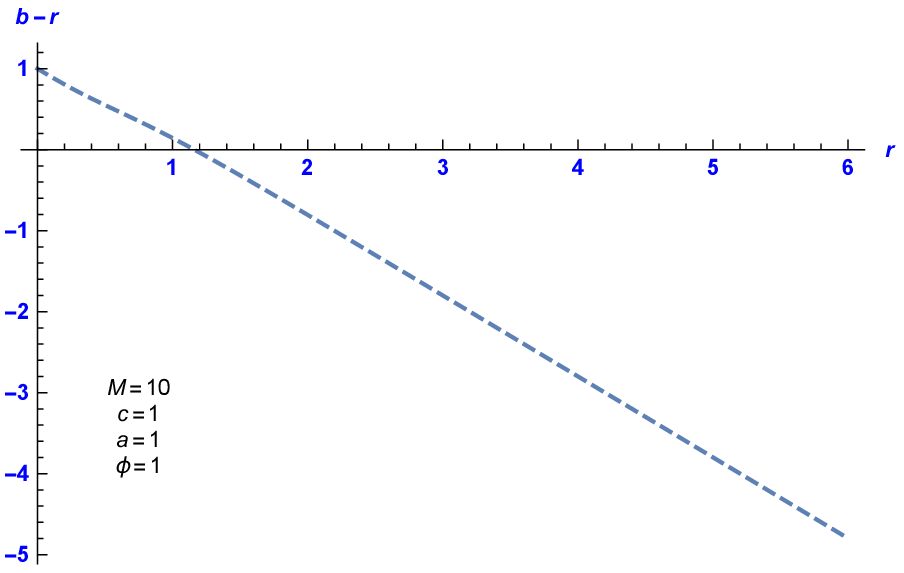}
 \label{26a}}\ \ \ \ \ \ \ \ \ \ \
 \subfigure[]{
 \includegraphics[height=4cm,width=4cm]{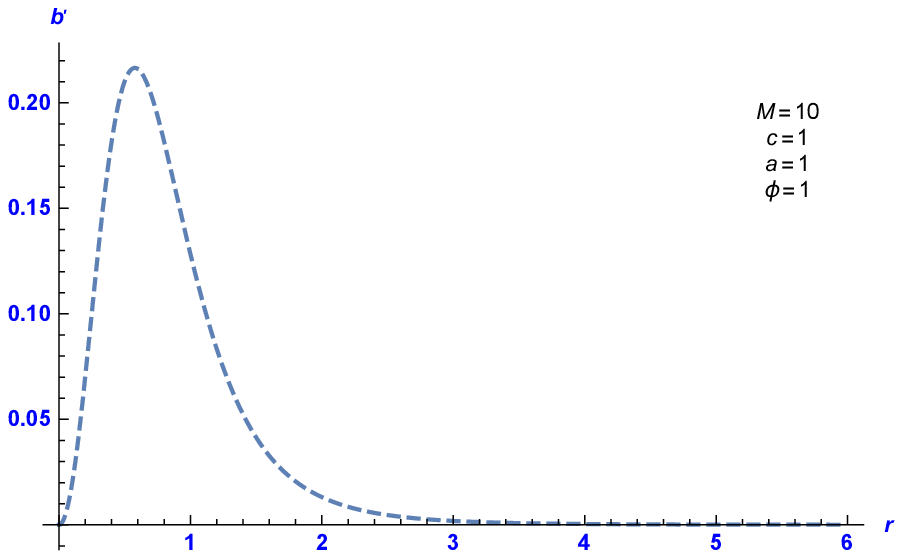}
 \label{26b}}
  \caption{\small{For $ n=\frac{1}{2}$. Left plot: $b-r$ versus $r$.  Right plot: $b'$ versus $r$. }}
 \end{center}
 \end{figure}

\begin{figure}[h!]
 \begin{center}
 \subfigure[]{
 \includegraphics[height=4cm,width=4cm]{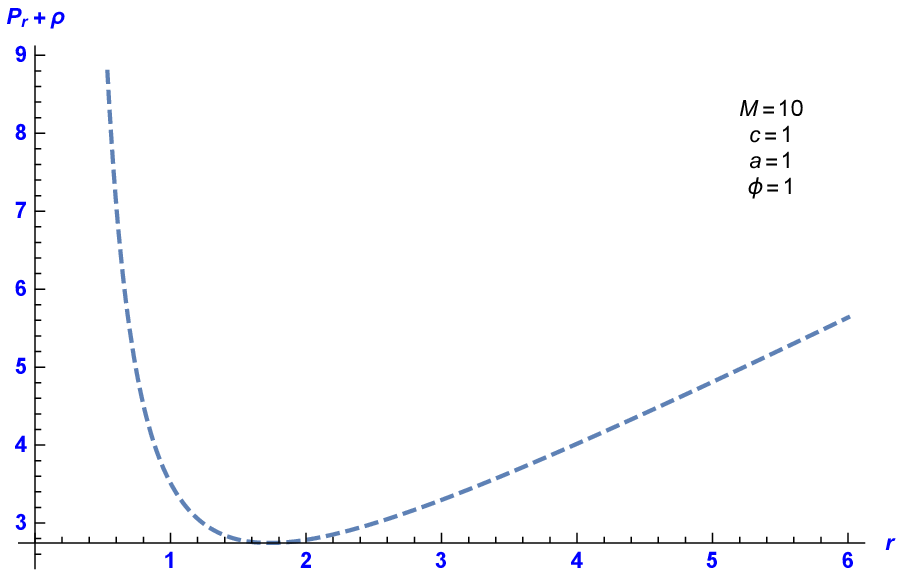}
 \label{27a}}\ \ \ \ \ \ \ \ \ \ \
 \subfigure[]{
 \includegraphics[height=4cm,width=4cm]{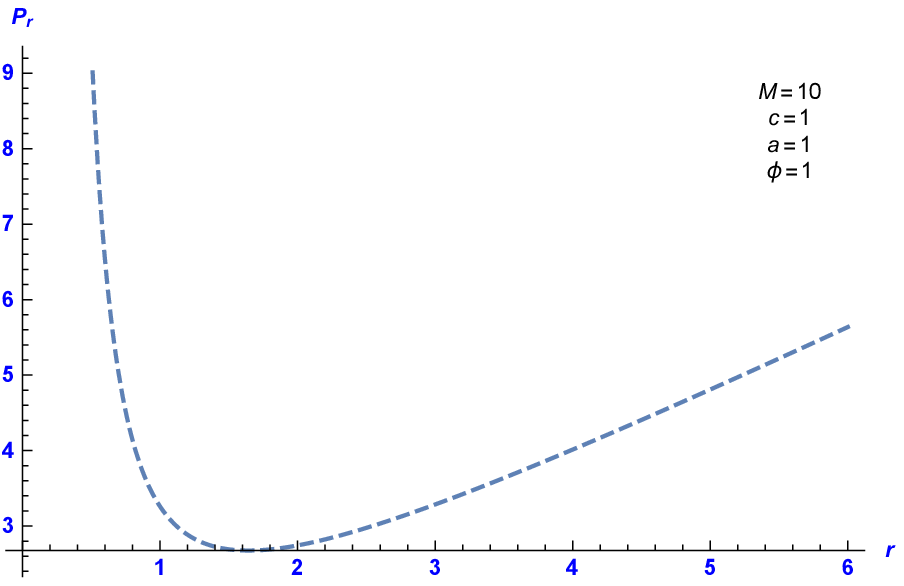}
 \label{27b}}
  \caption{\small{For $ n=\frac{1}{2}$. Left plot: $\rho+P_{r}$ versus $r$.  Right plot: $P_{r}$ versus $r$. }}
 \end{center}
 \end{figure}

We draw $\rho+P_{r}$ versus $r$ in Figs. \ref{27a}. The radial pressure plotted in Figs. \ref{27b}.  We see here that radial pressure is negative as well as $\rho+P_{r}$.\\

\begin{figure}[h!]
 \begin{center}
 \subfigure[]{
 \includegraphics[height=4cm,width=4cm]{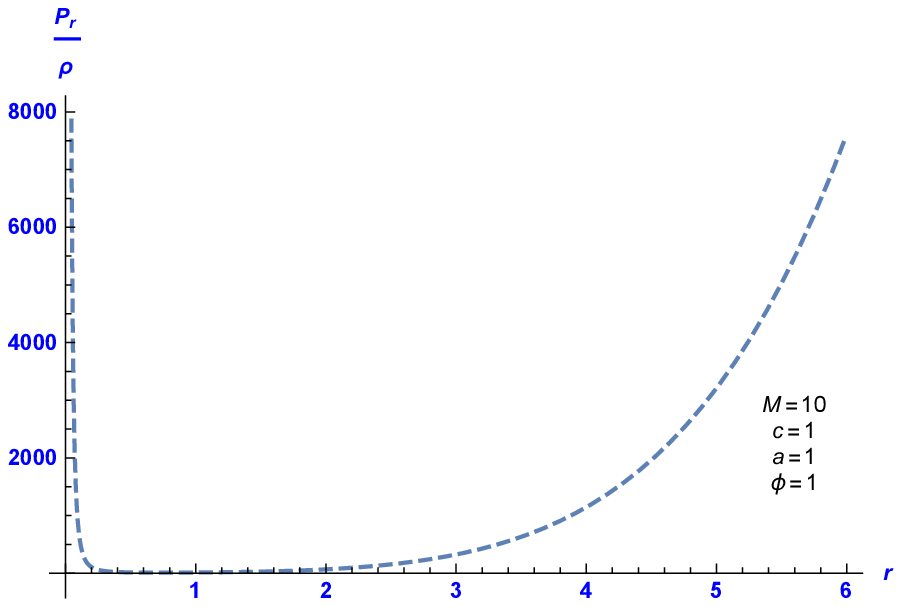}
 \label{28a}}
 \subfigure[]{
 \includegraphics[height=4cm,width=4cm]{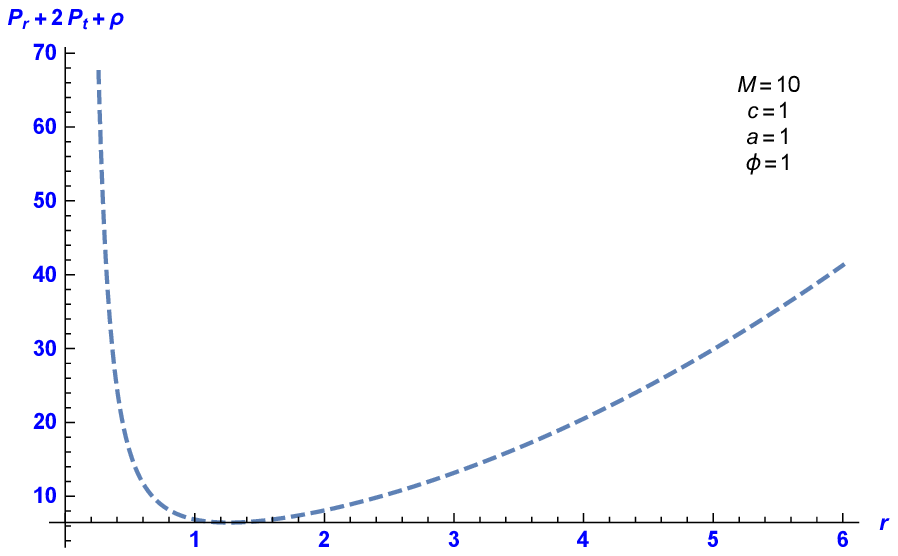}
 \label{28b}}
  \caption{\small{For $ n=\frac{1}{2}$. Left plot: $\frac{P_{r}}{\rho}$ versus $r$.  Right plot: $\rho+P_{r}+2P_{t}$ versus $r$. }}
 \end{center}
 \end{figure}

\begin{figure}[h!]
 \begin{center}
 \subfigure[]{
 \includegraphics[height=4cm,width=4cm]{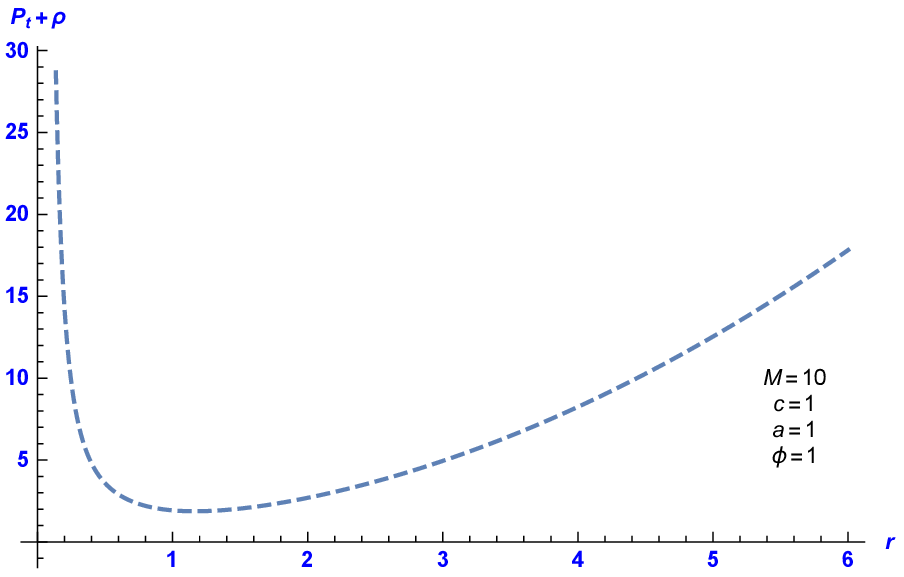}
 \label{29a}}
 \subfigure[]{
 \includegraphics[height=4cm,width=4cm]{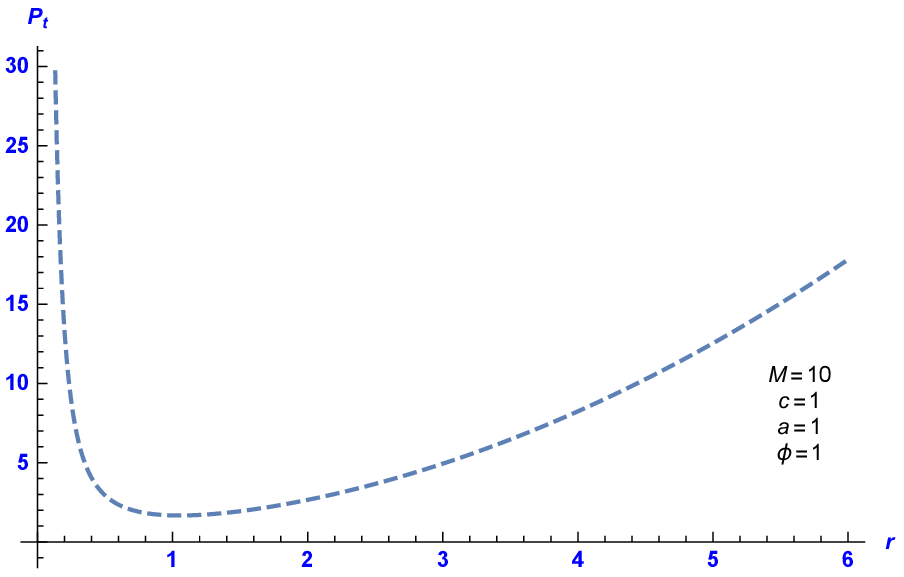}
 \label{29b}}
  \caption{\small{For $ n=\frac{1}{2}$. Left plot: $\rho+P_{t}$ versus $r$. Right plot:  $P_{t}$ versus $r$.}}
 \end{center}
 \end{figure}

The satisfaction and violation of the energy condition are depicted in  Figs. \ref{27a}, \ref{28b}  and  \ref{29a}. The transverse pressure is also plotted in Figs. \ref{29b}, to see that transverse pressure has both positive and negative values.\\

\begin{figure}[h!]
 \begin{center}
 \subfigure[]{
 \includegraphics[height=4cm,width=4cm]{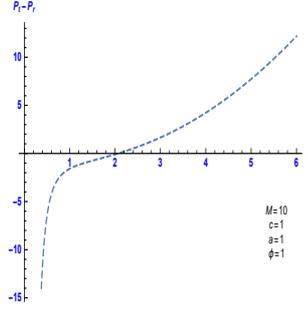}
 \label{30a}}
 \caption{For $ n=\frac{1}{2}$. Plot: $P_{t}-P_{r}$ versus $r$.}
 \end{center}
 \end{figure}

The behavior of equation of state and  anisotropy parameter  can be seen from Figs. \ref{28a} and \ref{30a}, respectively.\\

In the upcoming subsections, we shall study the other components of modified $f(R)$ gravity, namely,  $n=\frac{3}{2}$ and $n=\frac{1}{10}$.  One can see the effect of different value of $n$  on location of the wormhole throat and also radial pressure $P_{r}$ and transverse pressure ($P_{t}$) as well as the different energy conditions corresponding to NEC, WEC and SEC.

\subsection{Shape function, energy condition and equation of states for $n=\frac{9}{10}$}
\begin{figure}[h!]
 \begin{center}
 \subfigure[]{
 \includegraphics[height=4cm,width=4cm]{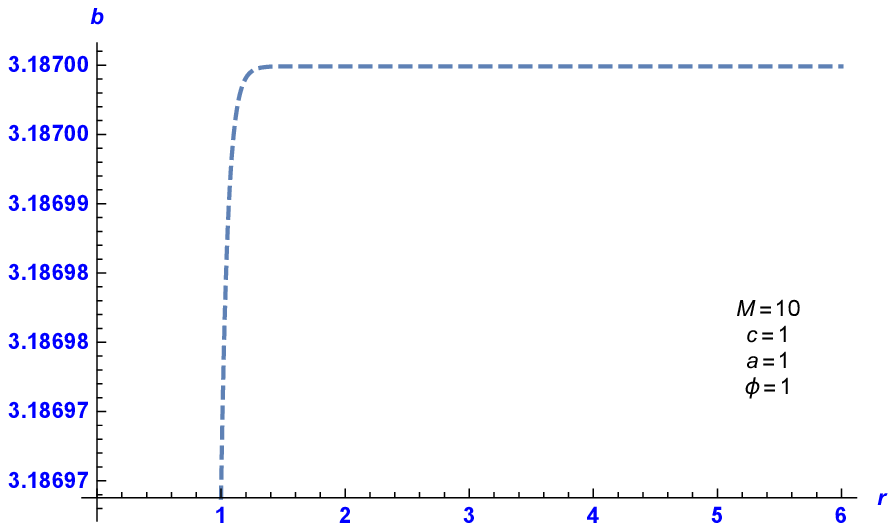}
 \label{31a}}\ \ \ \ \ \ \ \ \ \ \
 \subfigure[]{
 \includegraphics[height=4cm,width=4cm]{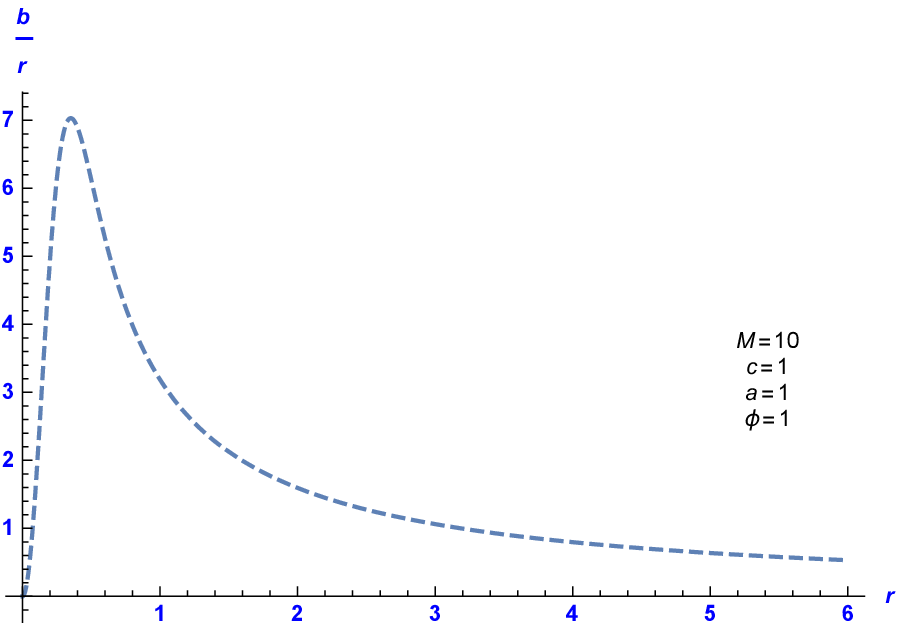}
 \label{31b}}
  \caption{\small{For $ n=\frac{9}{10}$. Left plot: $b$ versus $r$.  Right plot: $\frac{b}{r}$ versus $r$. }}
 \end{center}
 \end{figure}

In the case of  $n=\frac{9}{10}$, we obtain the following value of shape function:
\begin{equation}\label{31}
b(r)=1.94604\times10^{-8}r(\frac{M\sqrt{\phi}}{\alpha(\frac{1}{r^{2}})^{\frac{1}{10}}(\phi+r^{2})^{2}})^{10}(1+\frac{r^{2}}{\phi})^{20}\mathcal{H}\bigg[1.5,20,2.5,-\frac{1.1 r^{2}}{\phi}\bigg],
\end{equation}
where $\mathcal{H}$ is Hypergeometric $_{2}F_{1}$ function. Now, like the previous parts
 we study the energy condition for this model by plotting graphs.
 From Fig. \ref{31a}, one can see that
 the shape function   is an increasing function of  $r$ and takes infinitesimally small for
 small $r$. In Fig. \ref{31b}, we see that   $b/r$ takes only positive value and becomes asymptotically large when $r \rightarrow0$ as well as $r>1.5$.\\

\begin{figure}[h!]
 \begin{center}
 \subfigure[]{
 \includegraphics[height=4cm,width=4cm]{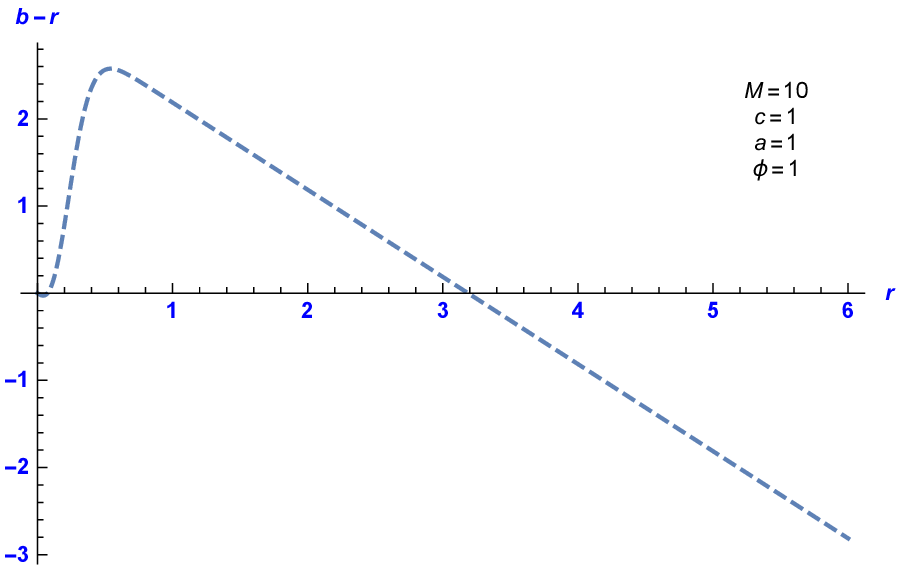}
 \label{32a}}\ \ \ \ \ \ \ \ \ \ \
 \subfigure[]{
 \includegraphics[height=4cm,width=4cm]{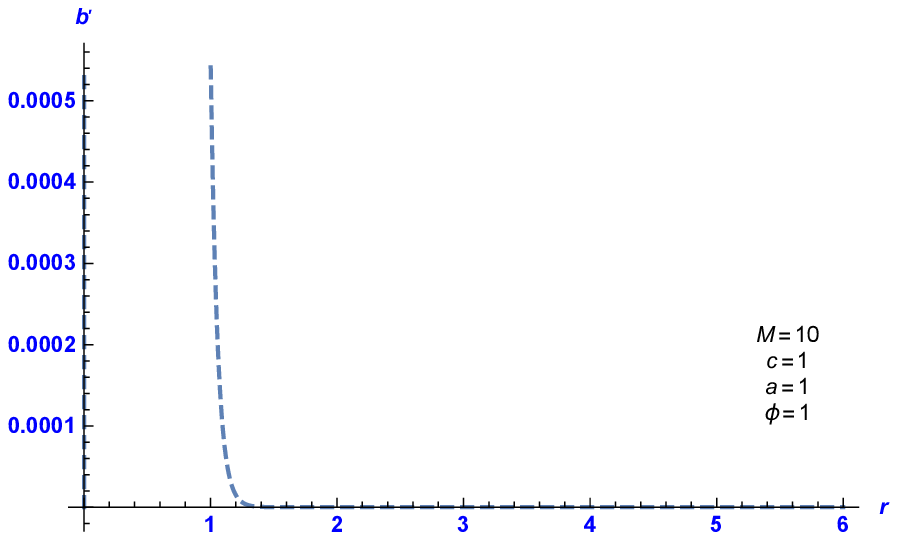}
 \label{32b}}
  \caption{\small{For $ n=\frac{9}{10}$. Left plot: $b-r$ versus $r$. Right plot:  $b'$ versus $r$. }}
 \end{center}
 \end{figure}

In Fig. \ref{32a}, we observe that $b-r$  cut the $r$-axis but approaches very closely
around $r=3$. Fig. \ref{32b}  justifies the violation of validity of the condition $b'(r_0) < 1$. This suggests that  the shape function does not satisfy all the requirements of warm hole structure.\\

\begin{figure}[h!]
 \begin{center}
 \subfigure[]{
 \includegraphics[height=4cm,width=4cm]{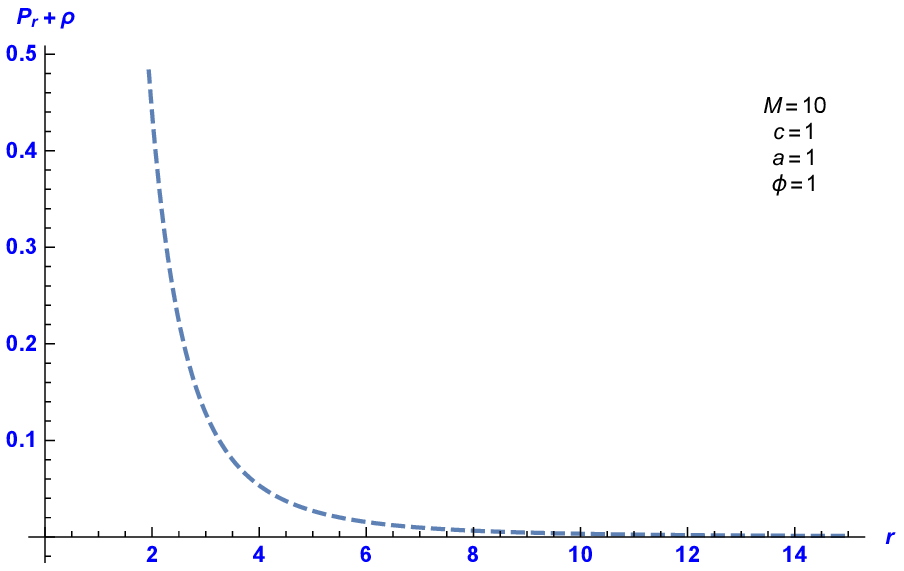}
 \label{33a}}
 \subfigure[]{
 \includegraphics[height=4cm,width=4cm]{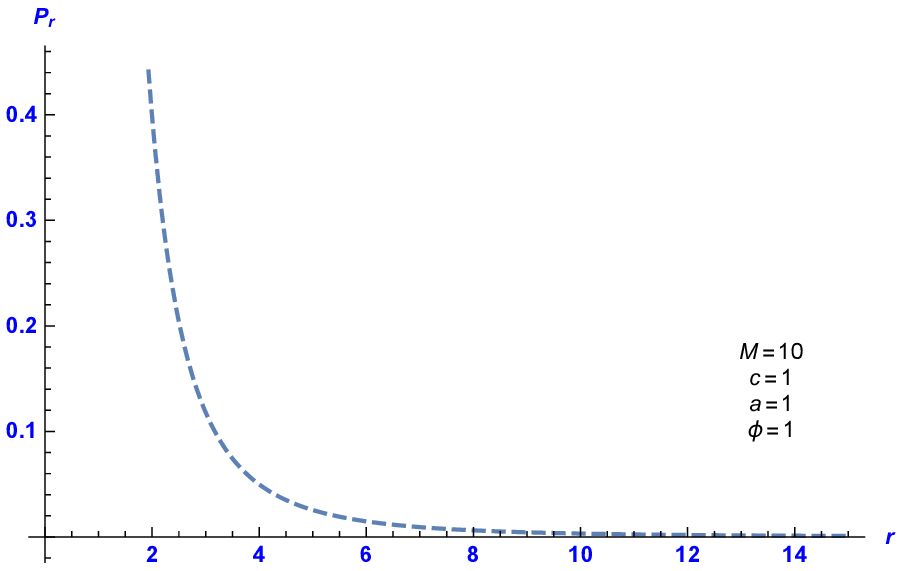}
 \label{33b}}
  \caption{\small{For $ n=\frac{9}{10}$. Left plot: $\rho+P_{r}$ versus $r$.  Right plot: $P_{r}$ versus $r$. }}
 \end{center}
 \end{figure}

\begin{figure}[h!]
 \begin{center}
 \subfigure[]{
 \includegraphics[height=4cm,width=4cm]{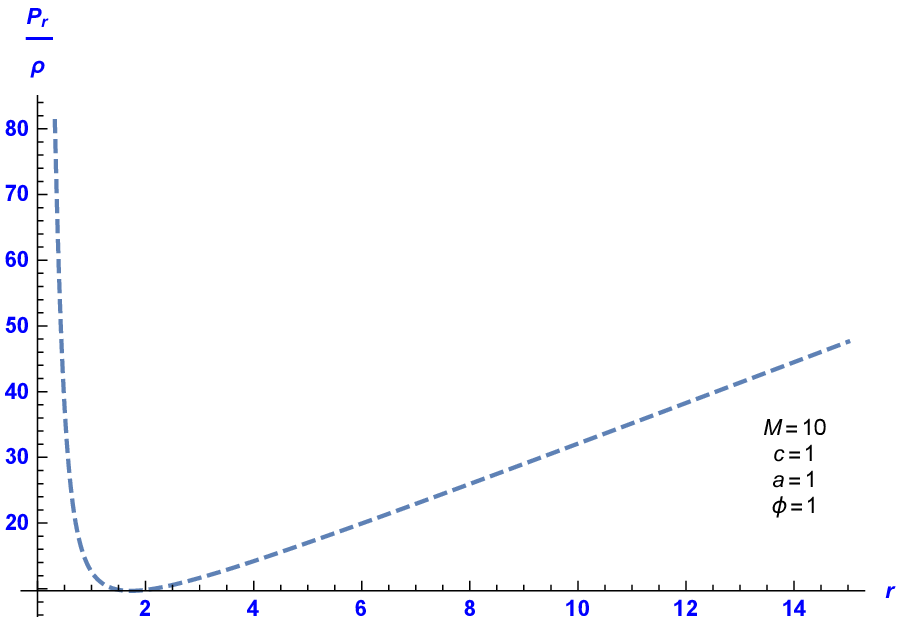}
 \label{34a}}
 \subfigure[]{
 \includegraphics[height=4cm,width=4cm]{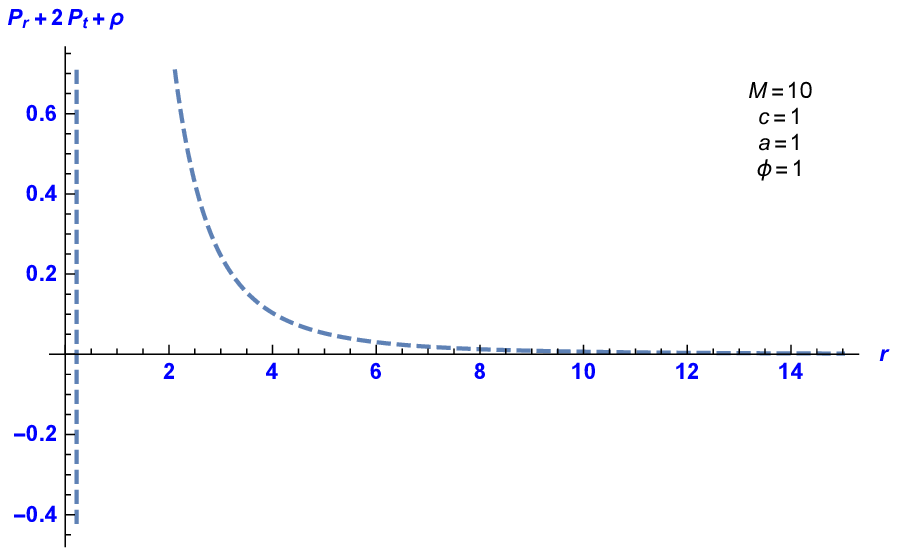}
 \label{34b}}
  \caption{\small {For $ n=\frac{9}{10}$. Left plot: $\frac{P_{r}}{\rho}$ versus $r$.  Right plot: $\rho+P_{r}+2P_{t}$ versus $r$. }}
 \end{center}
 \end{figure}

\begin{figure}[h!]
 \begin{center}
 \subfigure[]{
 \includegraphics[height=4cm,width=4cm]{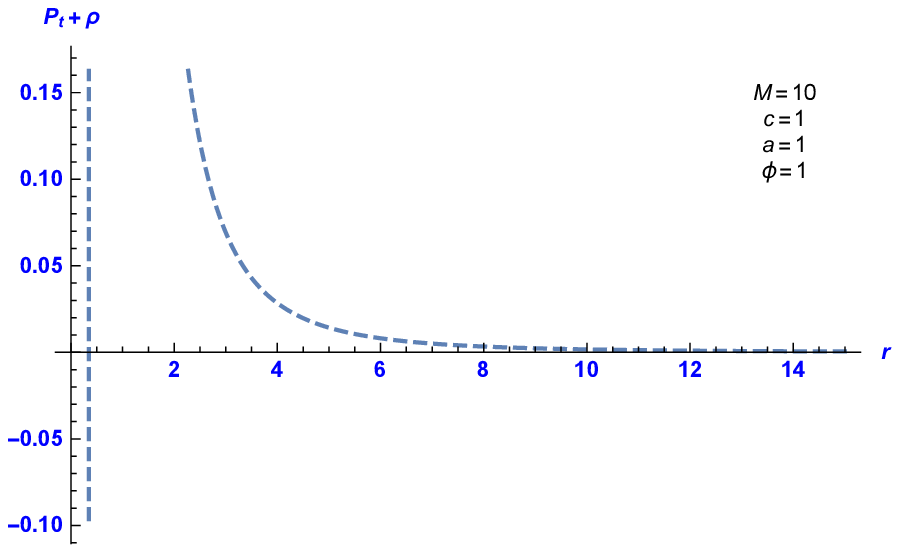}
 \label{35a}}\ \ \ \ \ \ \ \ \ \ \
 \subfigure[]{
 \includegraphics[height=4cm,width=4cm]{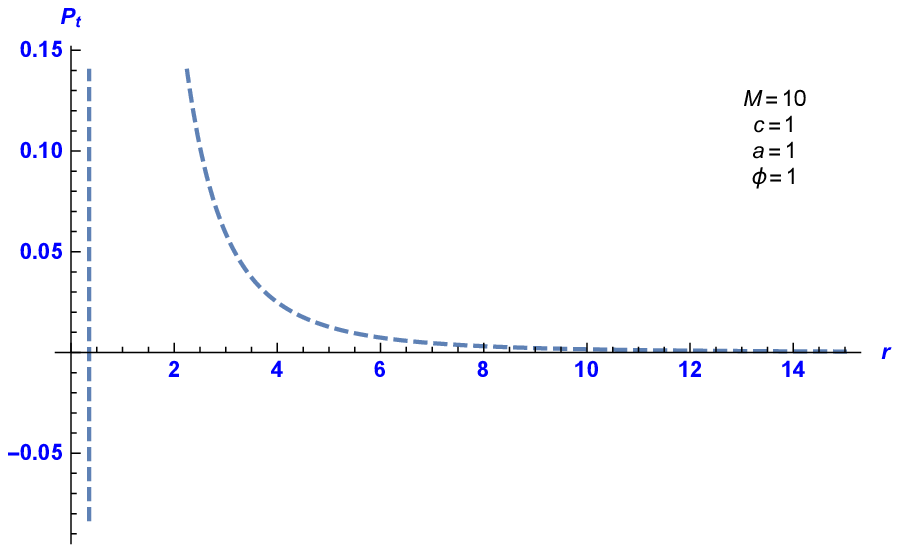}
 \label{35b}}
  \caption{\small{For $ n=\frac{9}{10}$. Left plot:  $\rho+P_{t}$ versus $r$. Right plot:  $P_{t}$ versus $r$. }}
 \end{center}
 \end{figure}

\begin{figure}[h!]
 \begin{center}
 \subfigure[]{
 \includegraphics[height=4cm,width=4cm]{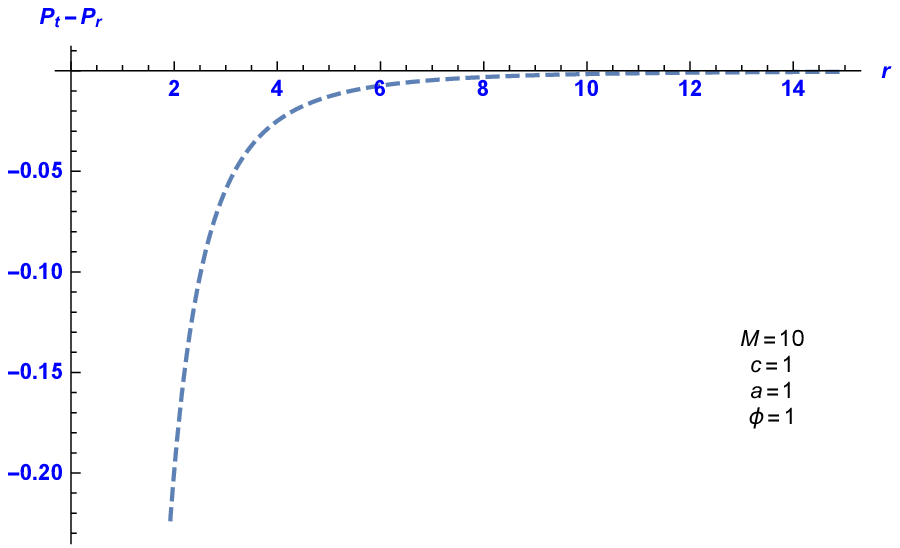}
 \label{36a}}
 \caption{For $ n=\frac{9}{10}$. Plot: $P_{t}-P_{r}$ versus $r$. }
 \end{center}
 \end{figure}

The radial pressure and transverse pressure are also plotted in figures \ref{33b} and  \ref{35b}, respectively. Here, we see  that the radial pressure  and transverse pressure
  both have asymptotically positive value when $r\rightarrow 0$. However, for large $r$
  both takes very small values but of opposite nature.
 The satisfaction and violation of different energy condition are depicted in  Figs. \ref{33a}, \ref{34b}  and  \ref{35a}. The
Fig. \ref{34a} suggests that  equation of state becomes negative for small values of $r$ and and takes asymptotically large value when $r\rightarrow 0$.  The  anisotropy parameter   is plotted in  \ref{36a}. The plot  shows that $P_t$ dominates over $P_r$.

\subsection{Shape function, energy condition and equation of states for $n=\frac{1}{10}$}
We start this subsection  by writing the shape function for the case of $f(R)$ gravity for  $n=0.1$
as follows
\begin{equation}\label{35}
b(r)=0.03r\left[\frac{M\sqrt{\phi}}{\alpha(\frac{1}{r^{2}})^{0.9}}(r^{2}+\phi)^{2}\right]^{1.1}\left(1+\frac{r^{2}}{\phi}\right)^{2.22} {}_2F_1\left[1.5,2.22,2.5,-\frac{r^{2}}{\phi}\right]+c.
\end{equation}
By plugging this value of shape function in (\ref{9.0}), one can obtained the curvature scalar ($R$).
It is difficult to solve the other equations numerically as the shape function contains hypergeometric function. So, we try to describe the results    by plotting some figures as following.

 \begin{figure}[h!]
 \begin{center}
 \subfigure[]{
 \includegraphics[height=4cm,width=4cm]{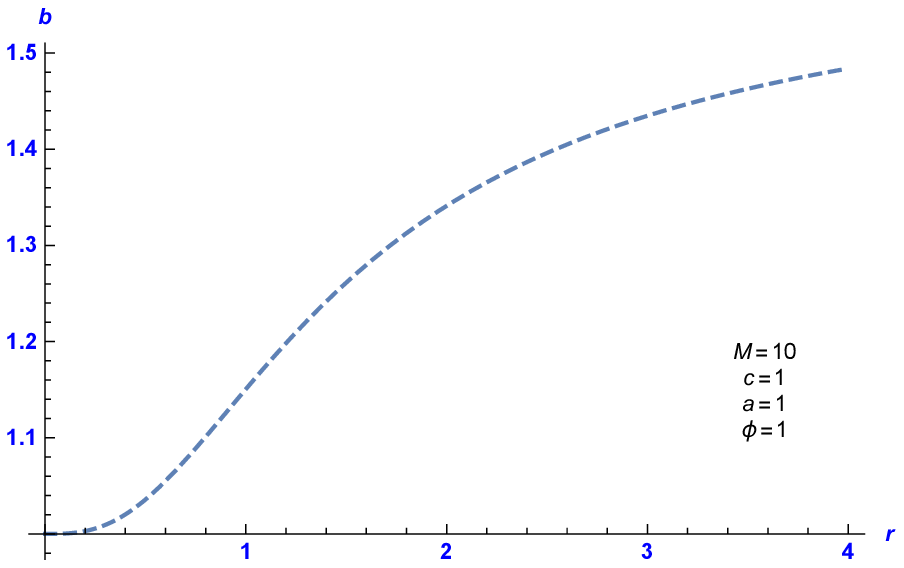}
 \label{37a}}\ \ \ \ \ \ \ \ \ \ \
 \subfigure[]{
 \includegraphics[height=4cm,width=4cm]{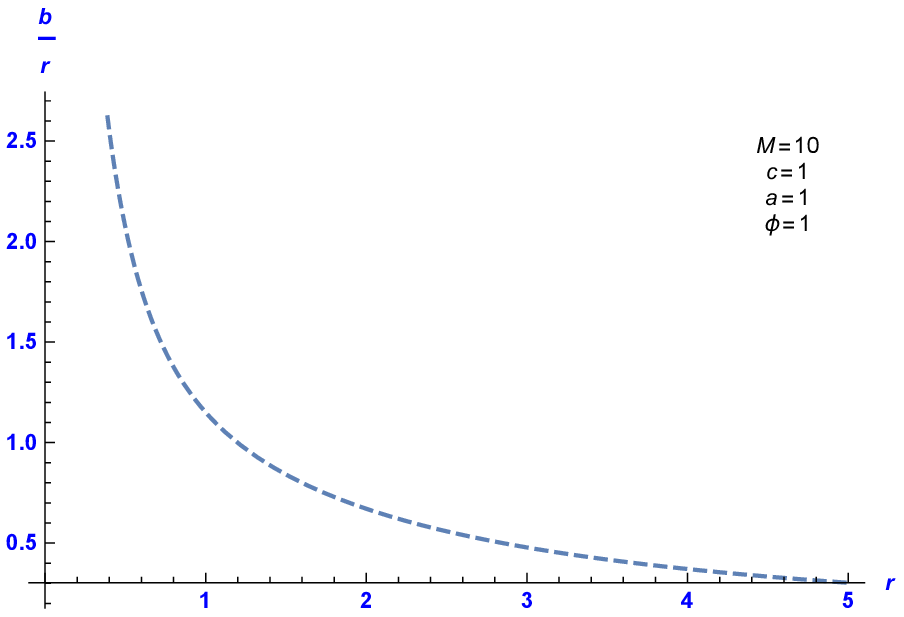}
 \label{37b}}
  \caption{\small{For $ n=\frac{1}{10}$. Left plot: $b$ versus $r$.  Right plot: $\frac{b}{r}$ versus $r$.}}
 \end{center}
 \end{figure}

 \begin{figure}[h!]
 \begin{center}
 \subfigure[]{
 \includegraphics[height=4cm,width=4cm]{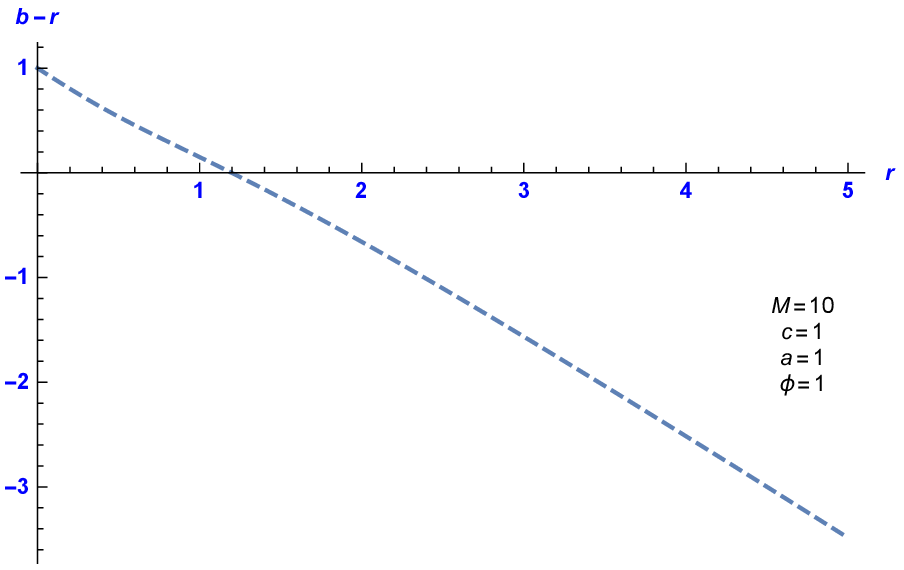}
 \label{38a}}\ \ \ \ \ \ \ \ \ \ \
 \subfigure[]{
 \includegraphics[height=4cm,width=4cm]{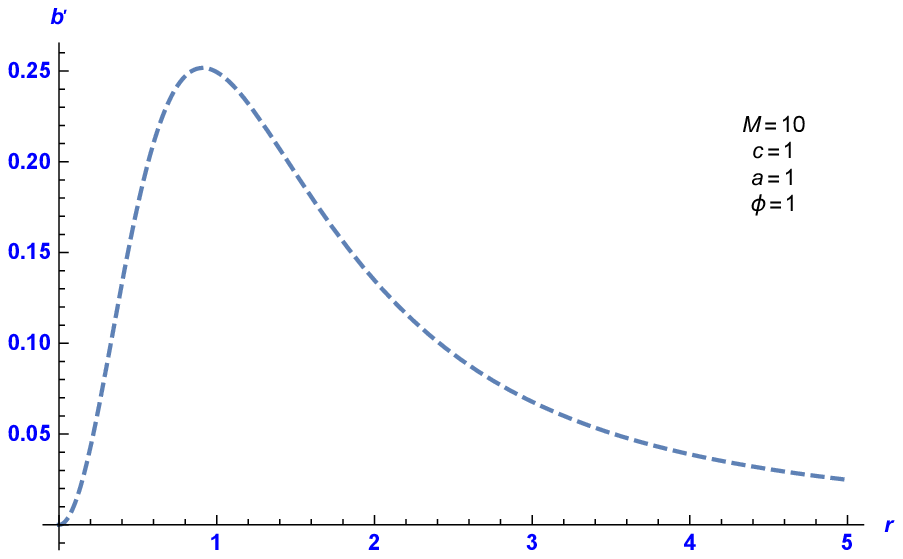}
 \label{38b}}
  \caption{\small{For $ n=\frac{1}{10}$. Left plot: $b-r$ versus $r$.  Right plot: $b'$ versus $r$. }}
 \end{center}
 \end{figure}

 \begin{figure}[h!]
 \begin{center}
 \subfigure[]{
 \includegraphics[height=4cm,width=4cm]{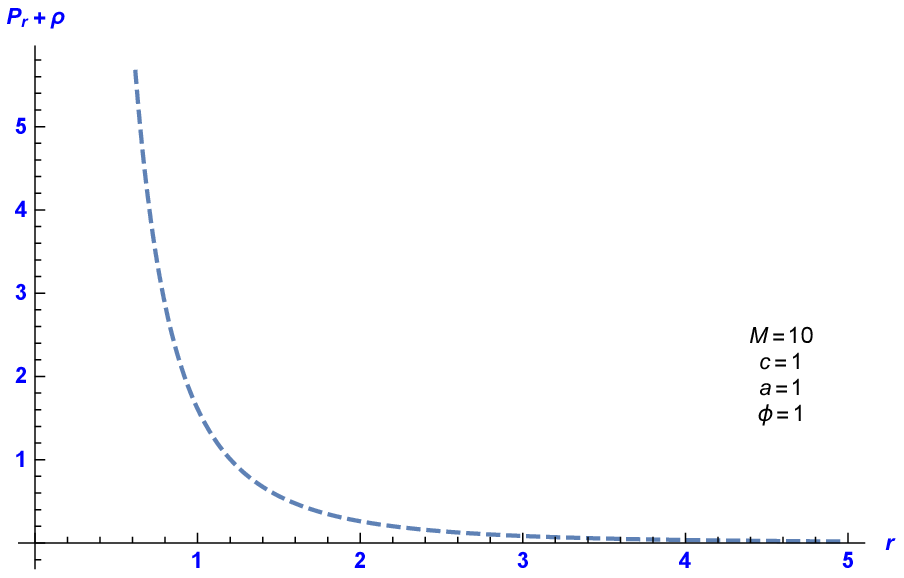}
 \label{39a}}\ \ \ \ \ \ \ \ \ \ \
 \subfigure[]{
 \includegraphics[height=4cm,width=4cm]{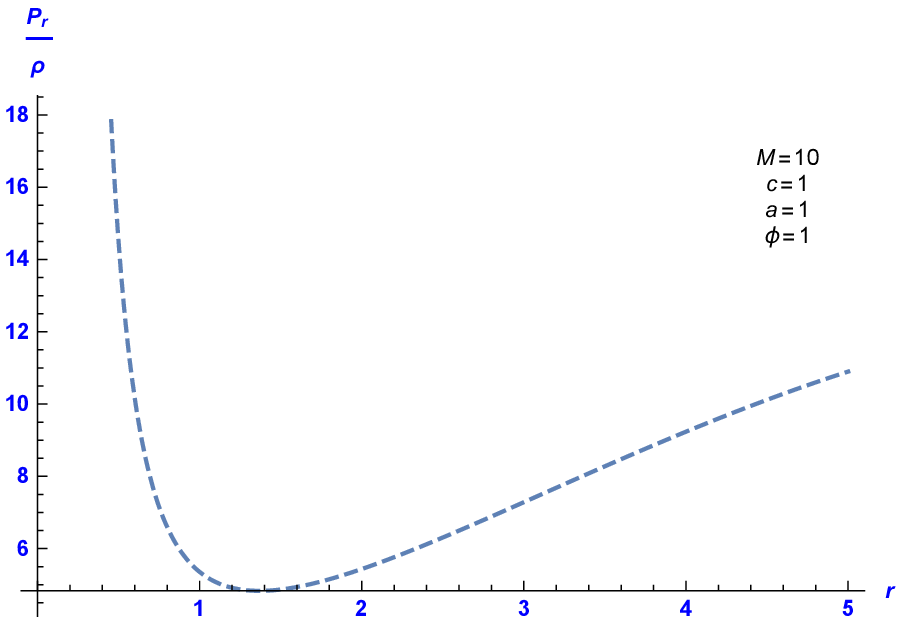}
 \label{39b}}
  \caption{\small{For $ n=\frac{1}{10}$. Left plot: $\rho+P_{r}$ versus $r$. Right plot: $\frac{P_{r}}{\rho}$ versus $r$. }}
 \end{center}
 \end{figure}

 \begin{figure}[h!]
 \begin{center}
 \subfigure[]{
 \includegraphics[height=4cm,width=4cm]{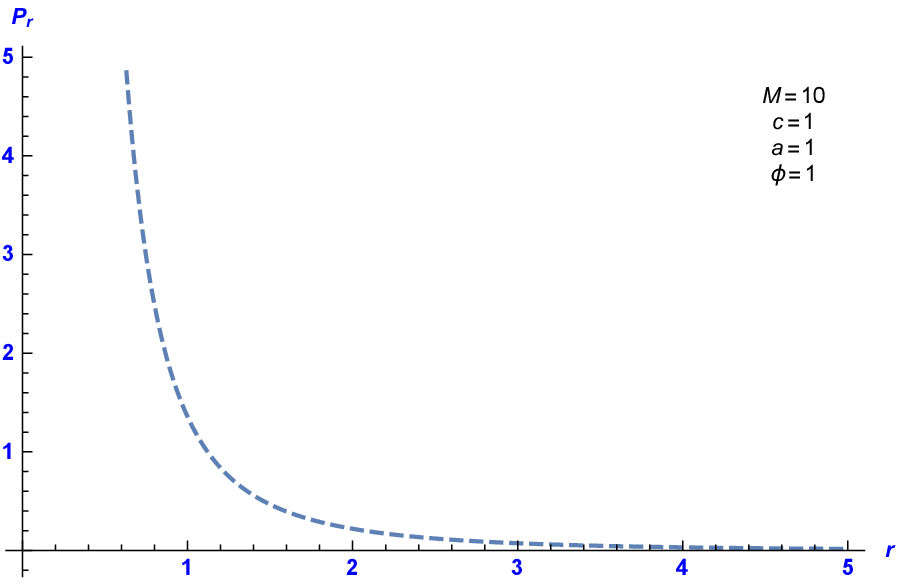}
 \label{40a}}
 \subfigure[]{
 \includegraphics[height=4cm,width=4cm]{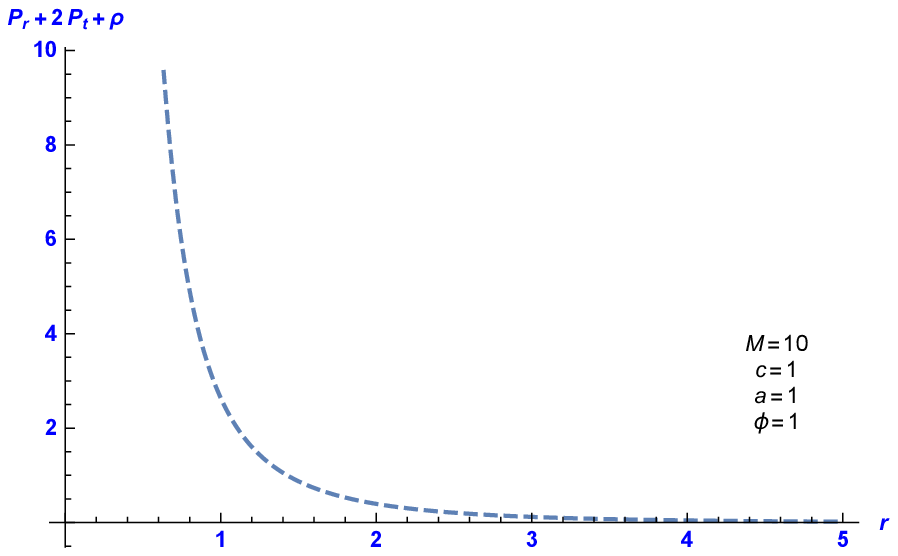}
 \label{40b}}
  \caption{\small{For $ n=\frac{1}{10}$. Left plot: $P_{r}$ versus $r$. Right plot: $\rho+P_{r}+2P_{t}$ versus $r$. }}
 \end{center}
 \end{figure}

 \begin{figure}[h!]
 \begin{center}
 \subfigure[]{
 \includegraphics[height=4cm,width=4cm]{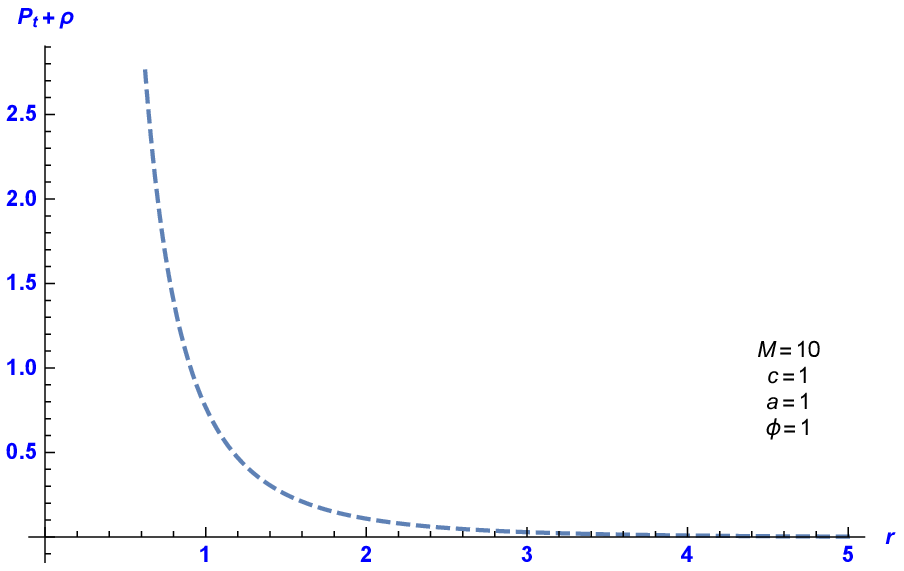}
 \label{41a}}
 \subfigure[]{
 \includegraphics[height=4cm,width=4cm]{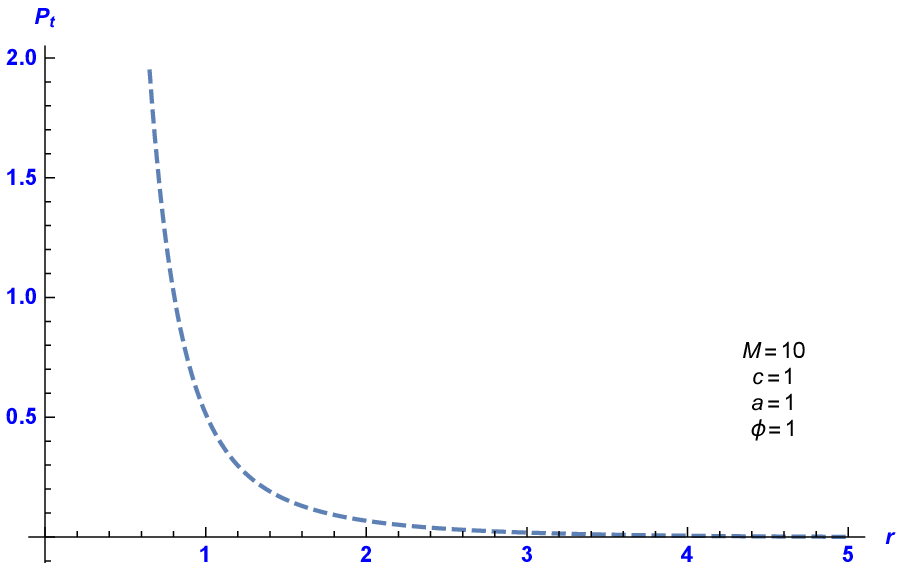}
 \label{41b}}
  \caption{\small{For $ n=\frac{1}{10}$. Left plot: $\rho+P_{t}$ versus $r$. Right plot:  $P_{t}$ versus $r$.}}
 \end{center}
 \end{figure}

 \begin{figure}[h!]
 \begin{center}
 \subfigure[]{
 \includegraphics[height=4cm,width=4cm]{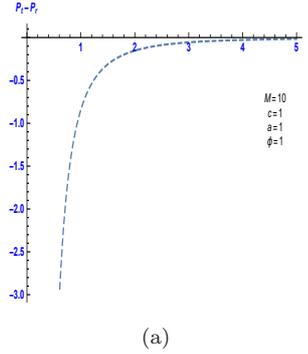}
 \label{42a}}
 \caption{For $ n=\frac{1}{10}$. Plot $P_{t}-P_{r}$ versus $r$.}
 \end{center}
 \end{figure}
From Fig. \ref{37a}, one can see that
 the shape function   is an increasing function of  $r$ and  and depends  on $r$ more or less linearly.  Fig. \ref{37b} confirms that asymptotically flat condition is being satisfied.
In the figure  \ref{38a}, we observe that $b-r$ curve   cuts the $r$-axis around $r=1.2$
which confirms the value of throat radius in this case.
From the Fig. \ref{38b}, we observe that the validity of the condition $b'
(r_0) < 1$ holds and therefore  the shape function  satisfies all the requirements of warm hole structure.
The radial pressure is  plotted in figure \ref{40a}, which confirms that the radial pressure takes asymptotically negative value when $r\rightarrow 0$ and takes zero value for large $r$.
The  transverse pressure is plotted in figure  \ref{41b}.  Plot suggests that transverse pressure is just
negative to that of radial pressure.
 The satisfaction and violation of different energy condition are depicted in  Figs. \ref{39a}, \ref{40b}  and  \ref{41a}. The  behavior of  equation of state can be seen from
Fig. \ref{39b}.  The  anisotropy parameter   is plotted in  \ref{42a} which shows that $P_t$ is a decreasing function of $r$ and  dominates over $P_r$. The important point in examining the two models in this paper is that problems arise for the energy condition for the first model in (n=0), i.e., SEC. Still, for certain values of r, energy conditions are fully established. Also, for the second model, the energy conditions in the range $0<n<1$ are well established, but when I consider ($n=0$), the wormhole throat and energy conditions have problems. Also, as we get closer to 1, both the wormhole throat and the energy conditions are accompanied by problems. From $0<n<1/2$, the energy conditions are favorable. When we are moving forward the values as $n>1/2$, the problems gradually occur, i.e., the energy conditions are no longer established for certain $r$.

At the end of this section, the energy conditions for different $n$ for this model can be briefly expressed in the following form.\\
$\bullet$ for $n=\frac{1}{2}$, $NEC$, $WEC$, and $SEC$ are satisfied.\\
$\bullet$ for $n=\frac{9}{10}$, $NEC$, $WEC$, and $SEC$ are satisfied in $n\gtrapprox 1$ .\\
$\bullet$ for $n=\frac{1}{10}$ , $NEC$, $WEC$, and $SEC$ are satisfied\\

We end up this section by commenting that different conditions corresponding to different energy conditions were examined by determining the shape function for the modified $f(R)$ gravitational model. The results are interesting as the energy condition and validity of shape function depends on the
value of $n$ in the particular choice of the model. So, the results can be used in searching
appropriate modified $f(R)$ model for wormhole.
\section{Concluding remarks}
It is known that wormholes are a series of imaginary objects whose geometry can solve Einstein equations by tolerating the violation of NEC. In the recent past, researchers have studied diverse wormholes according to different criteria, and they achieved elegance, bizarre, and exciting results that may be able to solve many mysteries. Keeping this in mind,  we have
investigated a series of exact solutions for a static wormhole following  smeared mass source geometry in the  modified $ f (R) $ gravitational model where Lorentzian distribution   resulting from a particle-like source has been considered. In particular, we have studied two particular models of modified $f(R)$
gravity model and have analyzed the results corresponding to  different cases  of these models.
We have first derived shape function for different values of $n$ in both the $f(R)$ models.
From the shape function, we have calculated the scalar curvature, radial pressure and transverse pressure.
Furthermore, we have studied the behavior of quantities by plotting the graph and validity of different energy condition, in particular, NEC, WEC and SEC. The energy of state and anisotropy parameter have also been
emphasized for each case. From the plots, we can estimate the value of throat radius for the wormhole.
 Remarkable, we have found that for $n=0$, there exist two shape functions showing different behavior.
 The results are really interesting which can be useful in exploring a suitable model for
 wormhole. It will be interesting to
 study further the wormhole under different conditions.
 Also, it is interesting to consider some corrections on $f(R)$ like logarithmic correction \cite{IJMPD} or use $f(R,L)$ theories \cite{PRD}.



\begin{thebibliography}{99}


\bibitem{1}
L. Flamm, Physik Z. 17, 448 (1916).
\bibitem{2}
A. Einstein and N. Rosen, Ann. Phys. 242, 2 (1935).
\bibitem{mult}
I. D. Novikov, A.A. Shatskiy and D. I. Novikov, [arXiv:1412.3749].
\bibitem{mult2}
J. Maldacena, A. Milekhin, F. Popov, [arXiv:1807.04726].
\bibitem{mult3}
J. L. Blazquez-Salcedo, C. Knoll, E. Radu, [arXiv:2010.07317].
\bibitem{3}E. Teo, Phys. Rev. D 58  024014 (1998).
\bibitem{4}P. K. F. Kuhfittig, Phys. Rev. D 67, 064015 (2003).
\bibitem{5}P. E. Kashargin and S.V. Sushkov, Grav. Cosmol. 14, 80  (2008).
\bibitem{6}M. Jamil, P. K. F. Kuhfittig, F. Rahaman and S. A Rakib, Eur. Phys. J. C 67, 513 (2010).
\bibitem{7} G. Clement, Phys. Rev. D 51  6803 (1995).
\bibitem{8}J. P. S. Lemos et al, Phys. Rev. D 68  064004 (2003).
\bibitem{9}F. Rahaman et al, Gen. Rel. Grav. 39  145 (2007).
\bibitem{10} J. P. S. Lemos and F. S. N. Lobo, Phys. Rev. D 78,  044030 (2008).
\bibitem{11}F. Rahaman et. al, Gen. Rel. Grav. 38,  1687 (2006).
\bibitem{12}F. Rahaman et. al, Mod. Phys. Lett. A24,  53 (2009).
\bibitem{13}F. Rahaman et. al, Class. Quant. Grav. 28,  155021 (2011).
\bibitem{14}F. Rahaman et .al, Int. J. Theor. Phys. 51,  1680(2012).
\bibitem{15} S.-W. Kim and H. Lee, Phys. Rev. D 63,  064014 (2001).
\bibitem{16} F. Rahaman et. al, Int. J. Theor. Phys. 48,  1637 (2009).
\bibitem{17} F. S. N. Lobo, Classical and Quantum Gravity Research, 1-78 (2008), Nova Sci. Pub.
\bibitem{18}M. S. Morris and K. S. Thorne, Am. J. Phys. 56,  395 (1988).
\bibitem{19}M. S. Morris et. al, Phys. Rev. Lett. 61,  1446 (1988).
\bibitem{20}T. A. Roman, Phys. Rev. D 47,  1370 (1993).
\bibitem{21}S. W. Hawking, Phys. Rev. D 46,  603(1992).
\bibitem{22}F. Rahaman, S. Sarkar, K. N. Singh and N. Pant, Mod. Phys. Lett. A 34,  1950010 (2019).
\bibitem{23} N. Godani and G. C. Samanta, Mod. Phys. Lett. A 34,  1950226 (2019).
\bibitem{24} N. Godani and G. C. Samanta,   New Astron. 80, 101399 (2020).
\bibitem{25} N. Godani and G. C. Samanta,   Eur. Phys. J.  C 80, 30 (2020).
\bibitem{26}G. C. Samanta and N. Godani,  Eur. Phys. J. C 79,  623 (2019).
\bibitem{27}O. L. Andino  and  C. L. Vasconez, [arxiv:2007.10422].
\bibitem{28} N. Sorokhaibam, [arxiv:2007.07169].
\bibitem{29} P. K. F. Kuhfittig, J. Appl. Math. and Phys. 8, 1263  (2020).
\bibitem{30} A. C. L. Santos, C. R. Muniz and L. T. Oliveira, [arxiv:2007.00227].
\bibitem{31}M. Chernicoff, E. Garcia, G. Giribet and E.  Rubin de Celis, JHEP 2020, 19 (2020) [arxiv:2006.07428].
\bibitem{32}C. X. Yan, T. Gansukh and Y. D.-han, [arxiv:2006.04344].
\bibitem{33} I. Fayyaz and M. F. Shamir, Eur. Phys. J. C 80, 430 (2020).
\bibitem{34}F. S. N. Lobo and  M. A. Oliveira, Phys. Rev. D 80, 104012 (2009).
\bibitem{35} S. H. Mehdipour, Eur. Phys. J. Plus 127, 80 (2012).; F. Rahaman et al., Phys. Rev. D 86,
106010 (2012); F. Rahaman et al., Phys. Lett. B 746, 73 (2015); F. Rahaman et al., Int. J. Theor. Phys. 54, 699
(2015).
\bibitem{CJP} M. Khurshudyan, B. Pourhassan, A. Pasqua, Can. J. Phys. 93, 449 (2015).
\bibitem{MNRAS} S. Capozziello, M. Faizal, M. Hameeda, B. Pourhassan, V. Salzano, S. Upadhyay, MNRAS 474, 2430 (2018)
\bibitem{36} O. Bertolami et. al,  Phys. Rev. D 75, 104016. (2007).
\bibitem{IJGMMP} M. Rostami, J. Sadeghi, S. Miraboutalebi, A. A. Masoudi, B. Pourhassan, Int. J. Geom. Meth. Mod. Phys.  17, 2050136 (2020).
\bibitem{PRD2} M., Cataldo, et al., Phys. Rev. D 79, 024005 (2009).
\bibitem{IJMPD}
J. Sadeghi, B. Pourhassan, A.S. Kubeka, M. Rostami, Int. J. Mod. Phys. D 25, 1650077 (2016).
\bibitem{PRD} B. Pourhassan and P. Rudra, Phys. Rev. D 101, 084057 (2020).
\bibitem{37}
Phongpichit Channuie, arXiv:1907.10605 (2019).
\bibitem{38}
L. Sebastiani, G. Cognola, R. Myrzakulov, S. D. Odintsov and S. Zerbini, Phys. Rev. D 89 2, 023518 (2014).
\bibitem{39}
R. Myrzakulov, S. Odintsov and L. Sebastiani, Phys. Rev. D 91 8, 083529 (2015).
\bibitem{40}
K. Bamba, R. Myrzakulov, S. D. Odintsov and L. Sebastiani, Phys. Rev. D 90 4, 043505 (2014).
\bibitem{41}
A. Codello, J. Joergensen, F. Sannino and O. Svendsen, JHEP 1502  050(2015).
\end{thebibliography}
\end{document}